\documentclass[letterpaper,12pt]{article}
\usepackage{tabularx} 
\usepackage{amsmath}  
\usepackage{amssymb}
\usepackage[margin=1in,letterpaper]{geometry} 
\usepackage{cite} 
\usepackage[final]{hyperref} 
\usepackage{float}
\usepackage{tablefootnote}
\usepackage{epstopdf}
\usepackage{placeins}
\usepackage{graphicx}
\usepackage{gensymb}
\graphicspath{ {main.tex} }
\usepackage{mathtools}

\usepackage[utf8]{inputenc}
\usepackage{setspace}
\usepackage{braket}
\usepackage[font=footnotesize,skip=4pt]{caption}
\usepackage{subcaption}
\usepackage[nice]{nicefrac}
\usepackage[ruled,vlined]{algorithm2e}
\usepackage{algpseudocode}
\usepackage{color,soul}
\usepackage{xcolor}

\begin{document}
\begin{titlepage}
   \begin{center}
       \vspace*{1cm}
       \huge
       Dynamics of Darwinian versus Baldwinian versus Lamarckian evolution

       \vspace{2.5cm}
       \small
       Kristoffer Reinholt Thomsen \& Steen Rasmussen

       \vspace{3.0 cm}
     
       \includegraphics[width=0.5\textwidth]{FLinT_logo_full_blue.png}
       \vspace{3.0 cm}\\
       FKF\\
       University of Southern Denmark\\
       April 2023
            
   \end{center}
\end{titlepage}







\newpage
\tableofcontents
\newpage
\linespread{1.25}

\setcounter{page}{1}

\begin{abstract}
    In recent years, studies in epigenetic inheritance in biological systems as well as studies on evolution in non-biological systems e.g., machine learning and robotics, have reopened the discussion of non-Darwinian methods of evolutionary optimization. In this paper, the three most prominent classical evolutionary strategies Lamarckian, Darwinian, and Baldwinian are implemented and compared in an agent-based simulation. The dynamics of optimization and learning are studied in constant as well as in dynamic environments, both as single evolutionary strategy populations and as mixed evolutionary strategy populations. The three different agent types are implemented as simple objects that need to gather resources to replicate. The agent fitness is defined through a bitstring match between the agent and the environment and to replicate each agent needs to gather sufficient resources in the simulated chemostat. Initially assuming constant replication costs for all strategies, it is not surprising that for single strategy environments the Darwinian strategy is superior (most quickly develops high fitness) for high learning costs, while the Lamarckian strategy dominates for low learning costs with the Baldwinian strategy placed in between. As expected, the Lamarckian strategy is also performing best in ‘fast’ varying environments, followed by the Baldwinian strategy with the Darwinian last. The opposite is true for ‘slowly’ varying environments, as the learning costs become a burden.  To make a more ’fair’ comparison between the three different evolutionary methods when they all interact, we can assume that both the ability to learn within an individual’s lifetime and the learning process itself require additional resources compared to an evolutionary process without lifetime learning.  When all three strategies coexist and we vary the replication costs for the different strategies we find the following general pattern: The evolutionary advantage decreases for Lamarckian and Baldwinian strategies as their learning costs increase and/or as their relative replication costs increase compared to the Darwinian strategies, and in most situations the Baldwinian strategy performs in between the Darwinian and Lamarckian strategies. Finally, as our investigations emphasize the superiority of the Lamarckian strategy in most cases with low learning cost, it is interesting to note that a corresponding biological inheritance mechanism did not evolve, while the Lamarckian strategy is currently extensively used in technology applications.
\end{abstract}

\section{Background} \label{introduction}

In this paper we use simulation to explore and compare evolutionary and learning strategies originally proposed by Lamarck, Darwin, and Baldwin. Because of the historical significance of the Lamarckian, Darwinian, and Baldwinian ideas, we start with a brief historical review in Section 1.1. In Section 1.2 we summarize some of the modern ideas on evolutionary and learning as well as compares and contrasts ideas from the past with perspectives for the future. In Section 2 we introduced the agent based simulation methods used, while Section 3 is devoted to the results obtained when comparing and contrasting Lamarckian, Darwinian, and Baldwinian evolutionary and learning strategies. Section 4 is devoted to a critical discussion of the results obtained, while Section 5 summarizes our findings.

\subsection{A Brief Historical Review of Lamarck's, Darwin's, and Baldwin's fundamental ideas}

\subsubsection{Lamarck}
Jean-Baptiste Lamarck (1744-1829) was a French naturalist and professor of zoology, who in his classical work \textit{Philosophie Zoologique} (1809) stated his formulation of a theory of biological evolution. Lamarck was a strong proponent of the naturalistic view of evolution and life as a process best explained by natural and universal laws indistinguishable from the laws governing the inorganic world \cite[pp.~8-9]{darwin1859origin}\cite[pp.~9-15]{lamarck18091984}. Through his study of zoology and paleontology, he also held a firm view of species being mutable in time, and all species, including man, as having descended from other species \cite[pp.~8-9]{darwin1859origin}.
Lamarck was a pioneer in the study of simple organisms, a field he felt was neglected by his peers, and he even coined the still-used classification term \textit{invertebrate} while working as "Curator of Insecta and Vermes" at the Muséum National d'Histoire Naturelle \cite{wiki_invertebrate}. Much of \textit{Philosophie Zoologique} is dedicated to explaining his philosophy behind classification of organisms. \par 

Lamarck is credited with formulating the first consistent theory of biological evolution which covered many levels of description in biology, ranging from simple organisms to cognition. However, he is mostly known for formulating two laws to which he attributed the modifications of the body structures of animals. He summarizes both laws as follows in  \cite[p.~113]{lamarck18091984}. The first law states:
\begin{quote}
    In every animal which has not passed the limits of its development, a more frequent and continuous use of any organ gradually strengthens, develops and enlarges that organ, and gives it a power proportional to the length of time it has been used; while the permanent disuse of any organ imperceptibly weakens and deteriorates it, and progressively diminishes its functional capacity, until it finally disappears.
\end{quote}
This first law is Lamarck's famous law of \textit{use and disuse}, which he saw as the channel of biological adaption. The second law states:
\begin{quote}
    All the acquisitions or losses wrought by nature on individuals, through the influence of the environment in which their race has been placed, and hence through the influence of the predominant use or permanent disuse of any organ; All these are preserved by reproduction to the new individuals which arise, provided that the acquired modifications are common to both sexes, or at least to the individuals which produced the young.
\end{quote}
The second law is what later has been condensed to the law of \textit{inheritance of acquired traits}. This is the inference for which Lamarck in modern times has been remembered, though not always fondly, since Lamarck's theory of inheritance was deemed incompatible after Mendel discovered genetic inheritance, and as science later discovered DNA as the molecular base for inheritance. Hereafter, the theory of Natural Selection proposed by Darwin progressively became the most supported theory of evolution among scientists.

In present times, however, Lamarck's ideas of evolution have again become relevant, both in biological and non-biological evolving systems. Epigenetic inheritance has reopened the discussion in biological systems, as such factors have been observed to allow transmission of different phenotypes from cells with identical genotypes \cite{jablonka1999epigenetic}. In non-biological systems, cultural evolution and evolving computational programs in the fields of artificial intelligence and artificial life can be seen as Lamarckian in nature, as evolution in such systems is often not solely driven by replication and random mutations. Furthermore, in non-biological evolution, replicating entities may be ill defined or the evolving entities may not even replicate in a Darwinian manner as we shall explore in the following sections. 


\subsubsection{Darwin}
\textit{The Origin of Species}, one of the most influential scientific writing ever written, was published in 1859 by Charles Darwin. As much has been written about Darwin's works, the following section is highly influenced by the English Wikipedia page on \textit{On the Origin of Species}, as this is possibly the most read text based on the original book, perhaps only second to the book itself.

Darwin included both evidence that he had collected on the Beagle expedition in the 1830s and his subsequent findings from research, correspondence, and experimentation. As historical context it should be noted that Charles Darwin's grandfather Erasmus Darwin outlined a hypothesis of transmutation of species in the 1790s; an idea Lamarck developed further in \textit{Philosophie Zoologique} \cite{lamarck18091984}. Both envisaged that spontaneous generation produced simple forms of life that progressively developed greater complexity, adapting to the environment by inheriting changes in adults caused by use or disuse. In his notes, Charles Darwin sketched a genealogical branching of a single evolutionary tree, discarding Lamarck's independent lineages progressing to higher forms. The central idea in Darwin's thinking is expressed in the following quote:
\begin{quote}
    As many more individuals of each species are born than can possibly survive; and as, consequently, there is a frequently recurring struggle for existence, it follows that any being, if it vary however slightly in any manner profitable to itself, under the complex and sometimes varying conditions of life, will have a better chance of surviving, and thus be naturally selected. From the strong principle of inheritance, any selected variety will tend to propagate its new and modified form. \cite[pp.~30-31]{darwin1859origin}
\end{quote}
In \textit{On the Origins of Species}, Darwin discusses the production of variation within species induced by selective breeding, and that this is central to understanding the natural world, where such variation is naturally, ubiquitously present \cite[chap. 1]{darwin1859origin}.

In Darwin's time there was no agreed-upon model of heredity. He accepted a version of the inheritance of acquired characteristics (which after Darwin's death came to be called Lamarckism), and in Chapter V of \textit{On the Origins of Species}, Darwin discusses what he called the effects of use and disuse; he wrote: \begin{quote}
    I think there can be little doubt that use in our domestic animals has strengthed and enlarged certain parts, and disuse diminished them; and that such modifications are inherited. Under free nature we have no standard of comparison by which to judge of the effects of long-continued use or disuse, for we know not the parent-forms; but many animals possess structures which can be best explained by effects of disuse. \cite[p.~235]{darwin1859origin}
\end{quote}
From this, it seems clear that Darwin did not think his theory of Natural Selection was a complete theory, and that phenotypic changes during the lifetimes of organisms were better explained by Lamarck's theory. Still, evolutionary research in the following centuries has been marked by competing Darwinian and Lamarckian factions. Building on Darwin's theory of Natural Selection, James Mark Baldwin later introduced a new factor, attempting to explain these ongoing phenotypic changes in organisms.



\subsubsection{Baldwin}
In his 1896 paper \textit{A New Factor in Evolution} \cite{10.23072453130}, the American philosopher and psychologist James Mark Baldwin argued that the Lamarckian idea of inheritance of acquired traits is redundant in evolution, as intelligence can explain the plasticity of the phenotype. Baldwin's main argument focused on the role of intelligence in organisms as a means of plastically adapting to their environments during their lifetimes, which gives the ability to increase survival and chance of reproduction. He was partly inspired by observing the natural mental development of his children, and his stepwise theory of cognitive development was a major influence on the later, and much more widely known, developmental theory of Jean Piaget \cite{wiki_james_mark_baldwin,cahan1984genetic}. Baldwin argued that: 
\begin{quote}
    The most plastic individuals would be preserved to do the advantageous things for which their variations show them to be the most fit, and the next generation would show an emphasis on just this direction in its variations \cite[p.~447]{10.23072453130}
\end{quote}

Baldwin's ideas were formulated within the Darwinian framework as he knew they had to be consistent with natural selection. Though he was hesitant to enter the field of theoretical biology, Baldwin proposed that his theory could explain discontinuities in fossil records \cite{10.23072453130}.

As the popularity of Lamarckism diminished, Baldwins ideas of phenotypic plasticity surfaced. A review of the role of phenotypic plasticity in evolution can be found in \cite{fusco2010phenotypic} by Fusco and Minelli, where phenotypic plasticity is defined as the ability of individual genotypes to produce different phenotypes when exposed to different environmental conditions. This includes the ability of an individual organism to change its phenotypic state or activity (e.g. its metabolism) in response to variations in environmental conditions. The current iteration of Baldwins new factor - the Baldwin effect - has been largely accepted as a present factor in evolution \cite{wiki_baldwin_effect}.




\subsection{A Brief Summary of Modern Evolutionary and Learning Studies}

\subsubsection{Evolution}
The molecular mechanisms of inheritance were still unknown in the time of Lamarck, Darwin, and Baldwin. After the rise of Mendelian genetics and the discovery of the distinction between the germ line and the somatic line, and later the discovery of DNA, Lamarck's theory was declared incompatible with biological evolution, as phenotypic changes during an organism's lifetime had no known means of being transmitted though inheritance \cite[p.~322]{dennett1996darwin}. Therefore, Darwin's idea of natural selection became the dominant theory of biological evolution. However, the study of epigenetics has since revived the fundamental ideas of Lamarckism in the field of biological evolution \cite{jablonka1999epigenetic}.


Modern theoretical and computational studies of biological evolution use multiple mathematical tools including game theory \cite{smith1979game,lindgren1992evolutionary}; systems of ordinary differential equations such as Manfred Eigen’s Quasi-species and Hypercycle models \cite{eigen1971selforganization,eigen1981transfer}; Stuart Kaufmann's NK model, which defines a rugged combinatorial fitness state space \cite{kauffman1987towards}; mathematical optimization, e.g., in John Holland’s Genetic Algorithms \cite{holland1992adaptation}; random graphs \cite{rasmussen1989toward}; cellular automata \cite{mccaskill2019analysing}; information theory \cite{wagner2017information}; statistical models \cite{fisher1926variability}; and agent based simulations 
\cite{6271f3a6f0274989aa6b0f632a87949b,farmer1986autocatalytic,kauffman1986autocatalytic,bagley1991artificial,fontana1993rna,tanaka2014structure,ofria2004avida}.
The Artificial Life and related communities have been deeply engaged in these developments.
See e.g., the Artificial Life and European Conference for Artificial Life Proceedings from the onset of conferences in 1987 (and proceedings in 1989) through today (2022).

Though evolution traditionally is mainly perceived as referring to biological evolution, the term is applicable to a much wider set of systems. The evolutionary framework is useful for describing any system in which entities are replicated or reproduced. For example, human made technologies also undergo evolutionary processes \cite{FLEMING20011019,solee2013evolutionary}. So far, mainly the evolution in physical technologies has been explored quantitatively. Physical technologies are the tools we use for transforming energy and materials in pursuit of our goals. However, social technologies - tools for organizing people in pursuit of our goals i.e., laws, moral values, money, religion, government, or business - also undergo evolution. Importantly, our physical and social technologies interact and undergo coevolution \cite{collaborators2020}. 

Cultural and technological evolution happens at a much shorter timescale than biological evolution, where human evolution occurs over thousands of years, while human culture, in particular human technology, may change over tens of years or even faster. Emergence of social and physical technologies such as the grammatical language, the written word, and the internet have increased the transmission scale, speed and accuracy between human agents over time. This ever-increasing communication and coordination capability is likely the main underlying driver of humanity’s increasing fitness over the last 200,000+ years \cite{sibani2020human}.

\subsubsection{Cognition and Learning}
It is obvious that an organism’s ability to react in response to environmental changes has a potential for enhancing its survival. Some single celled organisms have the ability to regulate the direction of their motility in response to concentration gradients and thereby move towards locations with more metabolic resources \cite{webre2003bacterial}. Therefore, intracellular information processing and simple decision-making already occur at the single cellular level. 

The neuron emerged as a specialized cell type capable of fast signal transmission both locally and across the entire organism. In some organisms a localized development of neuron-based information processing of sensory inputs eventually evolved into a central nervous system (CNS) and a brain. A defining feature of the brain is its capability of learning as a brain allows an organism to adaptively change its behaviors. The brain is plastic, as neural (and presumably intraneural) connections change during the lifetime of the organism. 

Can evolutionary processes be found in the neural networks of the brain? After all, the brain is a complex adaptive system that relies on optimization for continually and plastically improving the fitness of itself. A challenge for applying the evolutionary idea for describing the brain is the non-replicating neuron. In this context an interesting model using Darwinian channels in the brain has been proposed by Fernando et al. \cite{fernando2012selectionist} that instead considers multiple groups of neurons that can act in parallel. The model outlines a mechanism for copying patterns of synaptic connections between different neuronal groups, where one group forms connections to another group, and ”hijacks” it to make a copy of its synaptic topology. Thereby an existing neural connectivity can be copied (”replicated”) into a nearby neural group. With a sufficient number of such groups operating in parallel, the system composes a ”population” of neural groups, where the groups then define the units of selection. Thus, Darwinian-style natural selection might also occur in the brain \cite{fernando2012selectionist}.

It can be argued that the brain is fit for analysis using multiple (or even all) of the previously mentioned classical evolutionary theories; Lamarckism, Darwinism, and Baldwin’s new factor. The brain clearly exhibits Baldwinian characteristics, as its development, function, and microscopic structure, which are all phenotypic dimensions, are plastic. It is also clear that cultural evolution affects the human brain, as the behavior (which is generated by the brain) is highly influenced by verbal thoughts (thereby language), learned morals, and many other culturally transmitted phenomena. It has been argued that cultural evolution is partly Lamarckian \cite{dennett2017bacteria} making a full description of the brain partly Lamarckian, though this is not generally agreed upon \cite{hodgson2006dismantling}. Finally, it is clear from the evolutionary history of the development of the brain that genetic factors have a large influence on the structure of the brain \cite{madsen2019dr,godfrey2016other,jerison2012evolution}, making the evolution of the brain Darwinian. Darwinian mechanisms may also be relevant for describing neuroplasticity, as proposed by Fernando et al. (2012)  \cite{fernando2012selectionist}.

\subsubsection{Perspectives from the Past and the Future}

Historically, evolution has been associated with generational changes while learning has been associated with changes within the lifespan of an individual \cite{arita2000interactions}. This distinction makes sense as biological evolution has developed a material mechanism that ensures inheritance of traits from parent to the offspring based on genetic information. It should be noted that biological evolution apparently has not developed a material mechanism that ensures inheritance to offspring of behavior learned in the lifetime of a parent.

Biological evolution through Darwinian channels happens discretely through new material reproduction, whereas learning may happen continually throughout the entire lifespan of an organism through reorganization of an existing material basis within the organism. For higher order biological organisms this material reorganization happens within the brain, while for plants and some microorganisms this material reorganization occurs through distributed processes. In any event a hard distinction between evolution and learning is dissolved for systems undergoing Lamarckian and Baldwinian evolution, where adaption happens outside of replication/reproduction.

Evolution, biological as well as non-biological evolution, is sometimes an optimization processes, where traits become better and better suited to the environment, while qualitatively new traits do not emerge. However, qualitatively new traits do also emerge during evolution although it is not well understood how and under which conditions this occurs.  Therefore, current research on evolutionary processes is preoccupied with understanding the hallmarks \cite{bedau1991measurement,bedau1999visualizing} and mechanisms for open-endedness in such processes \cite{packardetal2019open}.
Thus, evolutionary processes come in two flavors \cite{rasmussen2019two}: optimization processes and expansion processes. The former suffices in systems whose size and interactions do not change substantially over time, while the latter is a key property of open-ended evolution, where new entities and interaction types enter the game. 
Examining data from systems in physics, biology, and engineering, the authors in \cite{rasmussen2019two} argue that the evolutionary optimization dynamics are the cumulative effect of multiple independent events, or quakes in these systems. These quakes are uniformly distributed on a logarithmic time scale and produce a decelerating fitness improvement when using the appropriate independent variable, which is a signature of evolutionary optimization. Finally, they argue that to enhance the evolutionary richness in a system that only optimizes, the system has to be enriched with new components and thereby interactions to expand its dynamics. 

A fundamental similarity between evolution and learning is that both can be characterized as processes that enhance outcomes over time. Further, evolution and learning are caused by similar underlying processes. They both fundamentally improve dynamics through trail and error. Even supervised machine learning is at the base a trail and error process that continuously is compared to a desired outcome. 

As our technologies become increasingly more life-like and intelligent, often implemented through Lamarckian and Baldwinian evolution, the lines between evolution and learning will likely further blend together. Take e.g. a self-driving car. The functionality/fitness of such a car is both dependent on its physical components (engine, tires, sensors, etc.) and the software which drives the car. The physical components cannot by updated - only exchanged with new identical versions for repair - whereas the software can be continually updated at any time. In this sense, the evolution of the physical car is Darwinian in nature, while the software is Lamarckian in nature. The environment of self-driving cars will also eventually be increasingly more dynamic as self-driving cars remove the constant and limiting factor of human driving capabilities, meaning future cars will optimise driving among other self-driving cars potentially allowing large and frequent innovations in transport. Note, the exact same logic of the self-driving car example can be applied to modern smartphones.

\newpage
\section{Methods} \label{Method}
To study the evolutionary dynamics and the role of learning in the three classical theories outlined in the previous section we develop an agent based model containing three agent types: Lamarckian, Darwinian, and Baldwinian agents. The capabilities of each agent type is inspired by their corresponding classical evolutionary theory. This model allows us to directly compare the fitness optimization and population dynamics of each classical theory. In this section, the model is described in detail, starting with the general characteristics shared by all three agent types.

\subsection{General agent characteristics}
\label{General agent characteristics}

Our simulation is composed of a fixed number of agents competing for space in a population of constant size $N$. The agents independently (no social interactions and no spacial dependency) collect resources from a reservoir at rates dictated only by their fitness (defined in Section \ref{Defining fitness and the environment}), and replicate after their total collected resources reach a threshold amount $R_{replication}$. During replication, the mother agent and daughter agent equally split the resources of the mother agent, as is typical with single-celled organisms. The daughter agent replaces the space of a random agent within the total population, thereby conserving the population size. Every agent carries information in the form of a genotype $g$ and a phenotype $p$ both defined as bit strings. The genome can be decomposed into two genes: a metabolic gene $M$ and a learning gene $Q$. The Darwinian genome only has the metabolic gene $M$ of length $|M| = L$, whereas the Lamarckian and Baldwinian in addition have the learning gene $Q$ also of length $|Q| = L$ that controls the learning rate, making their total genomes $2L$ long. Each agent in the initial population is ascribed a Family identifier (a number between 1 and $N$) which is inherited by descendent agents. These characteristics are shared by all three agent types.
\\
\\

\subsection{Environment and Fitness}\label{Defining fitness and the environment}
The environment is defined as a bit string $E$ of length $|E| = L$ that defines the optimal phenotype (bit string sequence) of the metabolic gene $M$. The normalized fitness of the $i$'th agent ($f_i$) is computed as the number of matching bits between the environment and the agent phenotype divided by the gene length: 
\begin{equation}
    \label{D_fitness}
    f_i^p(p_i,E)= \frac{\sum_{j=1}^L \delta_p}{L},\,\,\,\,\,\, 
    \textrm{where}\,\,\,\,\, 
    \delta_p = \begin{cases}
        1 \,\,\,\textrm{if}\,\,\,p_i(j)=E(j)\\
        0 \,\,\,\textrm{if}\,\,\,p_i(j)\neq E(j)\\
    \end{cases}
\end{equation}
i.e., the fitness function is linearly dependent on the number of matching bits. It also follows that the fitness landscape is smooth and contains a single global maximum corresponding to the phenotype with $L$ matching bits. The phenotypic fitness of an agent dictates the rate at which that agent gains resources:
\begin{equation}
    \label{general_resource_rate}
    r_i(t+\Delta t,f_i^p) = r_i(t,f_i^p) + f_i^p\Delta t.
\end{equation}
When all phenotype bits match the environment bits the agent has maximal fitness, and therefore also maximum resource gain per time step. Since the agents reproduce after having collected a threshold amount of resources, a high fitness also correspond to a high rate of replication.

In the code, the genotypic fitness $f_i^g$ is also computed to study the difference in dynamics of the genotype and phenotype. It is defined similarly to the phenotypic fitness:
\begin{equation}
    \label{D_fitness}
    f_i^g(g_i,E)= \frac{\sum_{j=1}^L \delta_g}{L},\,\,\,\,\,\, 
    \textrm{where}\,\,\,\,\, 
    \delta_g = \begin{cases}
        1 \,\,\,\textrm{if}\,\,\,g_i(j)=E(j)\\
        0 \,\,\,\textrm{if}\,\,\,g_i(j)\neq E(j)\\
    \end{cases}
\end{equation}
The pseudo code for computation of agent phenotypic fitness, genotypic fitness and resource gain is as follows: 
\\
\\


\begin{algorithm}[H]
\SetAlgoLined
\KwResult{Computation of phenotypic fitness, genotypic fitness and resource gain}
\For{$t=[1:T]$}{
 \For{$i=[1:N]$}{
  \For{$j=[1:L]$}{
   \If{$p_i(j)==E(j)$}{
    $f_i^p(t)=f_i^p(t)+1$\;
   }
   \If{$g_i(j)==E(j)$}{
    $f_i^g(t)=f_i^g(t)+1$\;
   }
   $r_i(t+1)=r_i(t)+f_i^p$\;
  }
 }
}

 \caption{Fitness and resource gain}
\end{algorithm}

Note that the both the genotypical and the phenotypical fitness can be calculated for all agent types, while it is only the phenotypical fitness that determines the ability to collect resources. 




\subsection{Darwinian agents}\label{section_Darwinian_agents}
 When reproducing, the Darwinian agent transmits the genetic information of the genotype bits with error rate $\alpha$. The genotype maps directly to the phenotype at birth and the Darwinian agent has no means of changing its phenotype during its lifetime. A Darwinian agent's genotype and phenotype are therefore equal and constant in time; $p_i^D = g_i^D$. This is a simplification as biological systems do not have a direct map from the genotype to the phenotype. Equation (\ref{D_fitness}) and (\ref{general_resource_rate}) therefore dictate that the Darwinian agent fitness is constant in time if the environment is also constant in time. The resource gain is modelled as follows:

\begin{equation}
     r_i(t+\Delta t,f_i) = 
     \begin{cases}
        \begin{aligned}
            &r_i(t,f_i^p) + f_i^p\Delta t \,\,\,&\textrm{if}\,\,\,\,\,\,r_i(t)<R^D_{replication}
            \\
            &\frac{r_i(t,f_i^p)}{2} \,\,\,&\textrm{if}\,\,\,\,\,\,r_i(t)\geq R^D_{replication}
            \\
        \end{aligned}
    \end{cases}
\end{equation}
The resource gain of the Darwinian agent is described by Equation (\ref{general_resource_rate}) when the agent has not yet reached the threshold required for reproducing. When the threshold is reached, the agent reproduces using half its current resources (as described in Section \ref{General agent characteristics}).

\begin{figure}[H] 
\centering
\includegraphics[scale=0.24]{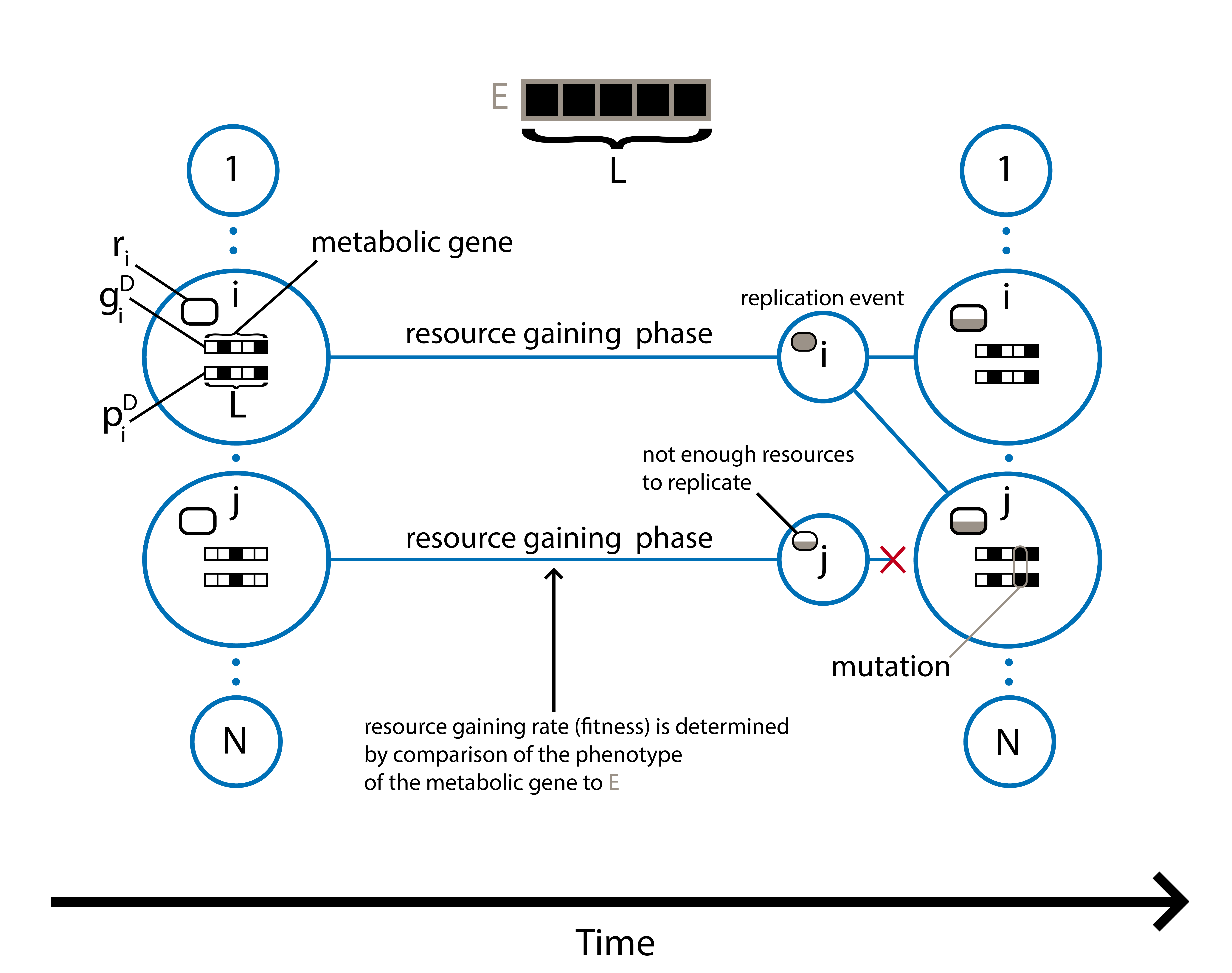}
\caption{Illustration of the Darwinian agent type. 
Note that $g^0_i = p^0_i$ and that the agent only has a metabolic gene $M$ of length $|M|=L$.
The agent gathers resources for a while until a certain threshold $R^D_{replication}$ is reached at which point agent $i$ divides. 
The remaining available resources are divided equally between the mother and the daughters agents, and a (set) of random genome mutations with frequency $\alpha$ may occur at the daughter genome.  
}
\label{replication_dyn}
\end{figure}

\subsection{Lamarckian agents}
\label{Lamarckian agents}


The defining feature of the Lamarckian agent is its ability to transmit its current phenotype at replication, i.e., the phenotype of the mother agent at the time of replication becomes the genotype of the daughter agent. The phenotype bit string is transmitted with no errors. The Lamarckian population instead evolves through individual agents changing their phenotypes during their lifetimes through so called learning events, in which agents switch the state of a random bit in their phenotypic information. Learning events occur with an agent-specific probability $c_i$ (denoted curiosity), and has a cost of a set amount of resources $R_{learning}$. This is an important assumption. Further, it is assumed that the material basis that enables the learning mechanism to exist requires extra resources to replicate, so $R^L_{replication} > R^D_{replication}$.

The rate at which Lamarckian agents gain resources is: 
\begin{equation} \label{L_resource_gain}
     r_i(t+\Delta t,f_i(t)) = 
     \begin{cases}
        \begin{aligned}
            &r_i(t,f_i^p(t)) + \left[f_i^p(t)-R^L_{learning} c_i\right]\Delta t \,\,\,&\textrm{if}\,\,\,\,\,\,r_i(t)<R^L_{replication}
            \\
            &\frac{r_i(t,f_i^p(t))}{2} \,\,\,&\textrm{if}\,\,\,\,\,\,r_i(t)\geq R_{replication}
            \\
        \end{aligned}
    \end{cases}
\end{equation}
Curiosity is a genetic trait added to the genome as a second gene $Q$ also of length $L$, making the total Lamarckian genome a length of $2L$ bits. The value of the $i$'th agent's curiosity is calculated as follows 
\begin{equation} \label{curiosity}
    c_i = \frac{\sum_{j=L+1}^{2L} p_i(j)}{L}.
\end{equation}
This yields $L$ possible values: 
\begin{equation} \label{c_states}
    c_i \in C\cdot \left\{\,\,0,\,\,\frac{1}{L},\,\,\frac{2}{L},\,\,\cdots\,\,,\,\,\frac{L-2}{L},\,\,\frac{L-1}{L},\,\,1\,\,\right\},
\end{equation}
where $0\leq C \leq 1$. $C$ is a parameter defining the maximum curiosity of agents. Learning events may change any bit in the phenotype; both in the metabolic gene (from which the fitness is computed) and in the curiosity gene. 

For the agent to be able to effectively learn from learning events in the phenotypical gene, the agent memorizes the location of the previously switched bit and its original state, as well as its resource gain rate during the previous time step. If the agent has decreased its resource gain rate after the bit switch, it switches the learning bit back, reversing its phenotype to its original state before learning. This is described in the Algorithm 2 pseudo code. Note that the switched bit may be within either of the two genes (metabolic and curiosity). If a learning event change the curiosity gene, the fitness of the agent is not changed (in a constant environment). Agents will therefore not reverse switches in the curiosity gene.
\\
\\

\begin{algorithm}[H]
\SetAlgoLined
\KwResult{Changes in phenotype due to learning}
 \For{$t=[1:T]$}{
  \For{$i=[1:N]$}{
  $learning\_bit$ = random bit between $1$ and $2L$\;
  $p_i(learning\_bit)$ = not($p_i(learning\_bit)$)\;
  Compute and store $r_i(t+1)$ using Equation \ref{L_resource_gain}\;
  $\Delta r_i(t+1) = r_i(t+1)-r_i(t)$\;
   \If{$\Delta r_i(t+1)<\Delta r_i(t)$}{
   $p_i(learning\_bit)=$not($p_i(learning\_bit)$);
  }
 }
}
 \caption{Learning algorithm}
\end{algorithm}
\vspace*{-10mm}

\begin{figure}[H] 
\centering
\includegraphics[scale=0.24]{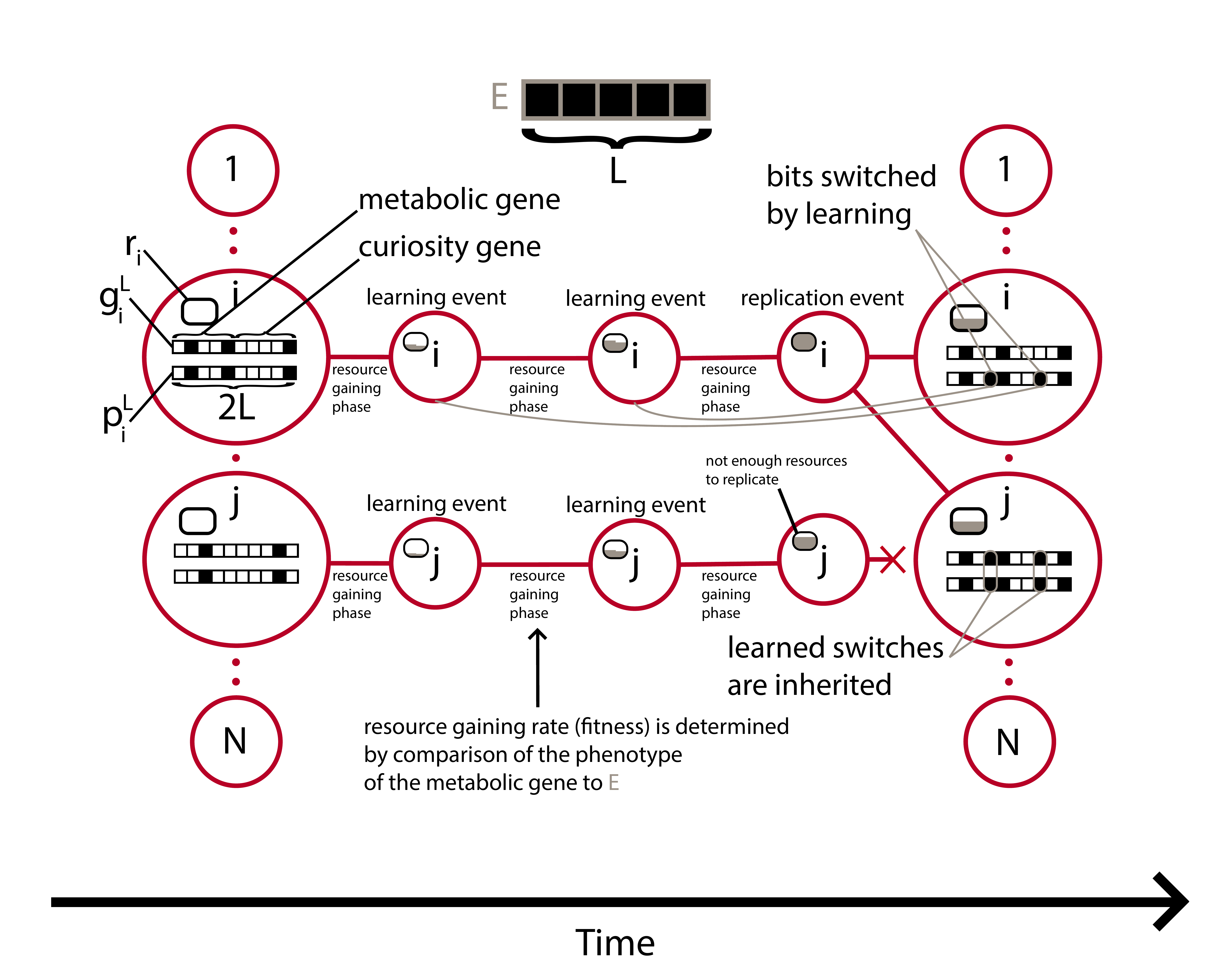}
\caption{Illustration of the Lamarckian agent type.
Note that the agent has two genes each of length $L$: one for metabolism and one for learning. 
The agent gathers resources while it at the same time, with a frequency determined by the curiosity parameter $C$, learns about its environment. 
Each learning event requires a certain amount of resources $R_{learning}$ and the learning process can occur at both phenotypical genes, so the agent can both alter its metabolic abilities as well as it 'curiosity' level during it's lifetime.
At a certain resource threshold $R^L_{replication}$ the agent $i$ divides and its resources are divided equally between the mother and the daughter agents.
Note that the phenotypical genes are inherited, not the genotypical genes.}
\label{replication_dyn}
\end{figure}

\subsection{Baldwinian agents} \label{sec: B method}
The Baldwinian agent may be viewed as the missing link between the Darwinian agent and the Lamarckian agent. Like the Darwinian agent, the Baldwinian agent only transmits the genotype with error rate $\alpha$ during replication. However, like the Lamarckian agent the Baldwinian agent is capable of learning during its lifetime, changing its full phenotype (both $M$ and $Q$ genes) without being able to transmit these changes to its offspring. The ability to learn via the curiosity trait can be seen as modelling phenotypic plasticity, as Baldwinian (and Lamarckian) agents during their lifetime $\tau_i$ explore approximately $c_i\cdot \tau_i$ new phenotypic states. The curiosity and resource gain (fitness) of the Baldwinian agent is defined by Equations (\ref{L_resource_gain}), (\ref{curiosity}) and (\ref{c_states}), like the Lamarckian agent type. If $C=0$ in Equation (\ref{c_states}), the Baldwinian agent type coincides with the Darwinian agent type.

Note that in the current implementation the Baldwinian agent type does not mutate its curiosity gene during replication. A Baldwinian agent may change its phenotypic curiosity during its lifetime through learning, but will not transmit such changes to its daughter agents. Thus, a population of Baldwinian agents have a limited pool of genotypic curiosity; meaning Baldwinian populations will eventually have a homogeneous genotypic curiosity after sufficiently many generations, though there may still be variations in their phenotypic curiosity (if the genotypic curiosity of the population is non-zero).

\begin{figure}[H] 
\centering
\includegraphics[scale=0.24]{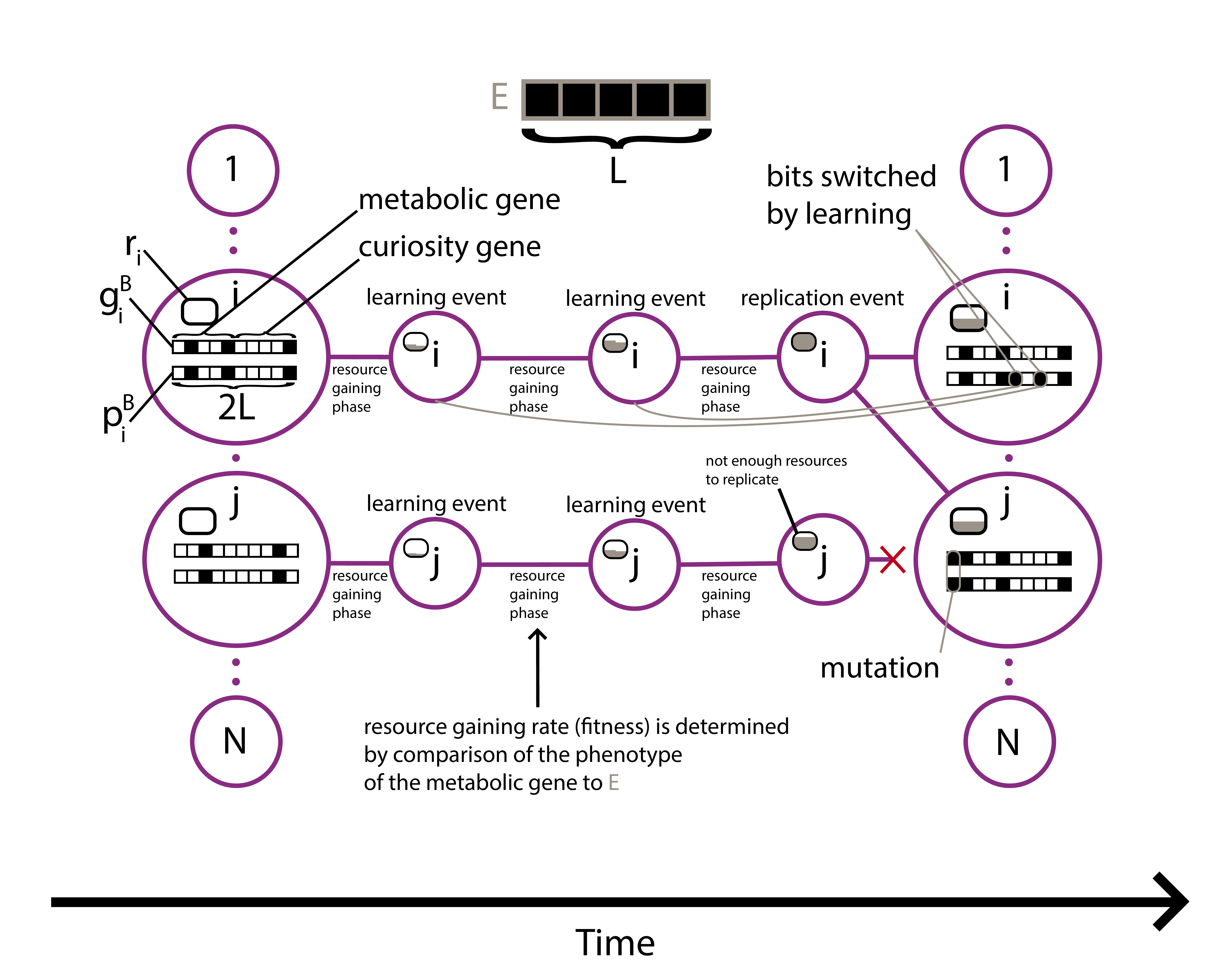}
\caption{Illustration of the Baldwinian agent type. Like the Lamarckian agent type, the Baldwinian agent type has both a metabolic and curiosity gene, and can change the phenotype of both genes via learning events at the cost of $R_{learning}$. The illustration shows agent $i$ gathering resources and learning until it has reached a certain threshold $R^{B}_{replication}$ where it replicates; its daughter taking agent $j$'s place in the population. However, learned phenotypic changes are not transmitted during replication, where the genome instead is mutated, as for the Darwinian agent type.}
\label{replication_dyn}
\end{figure}

\subsection{Temporally dependent Environment} \label{sec: E(t)}
The environment $E$ is previously defined by a bit string that is constant in time, usually as a string of length $L$ containing only ones. However, the environment may also be time dependent. How the environment is a function of time may be defined in numerous ways, giving rise to different types of dynamics. In our model, the time varying environment is defined as a semi-periodic variation, such that time periods of length $\tau_E$ with constant environment, is interrupted by sudden environmental changes given by an environmental mutation (error) rate $\alpha_E$ (See Algorithm 3).
This loosely models the biological emergence of new feeding conditions in a chemostat.
\\
\\

\begin{algorithm}[H]
\SetAlgoLined
\KwResult{Periodic mutations in the environment}
 \For{$t=[1:T]$}{
  \If{$remainder(t,\tau_E)==0$}{
   \For{$j=[1:L]$}{
    Draw $random\_number$\; 
    \If{$random\_number<\alpha_E$}{
    $E(j)$=not$(E(j))$
    }
   }
  }
 }

 \caption{Temporal changes in the environment}
\end{algorithm}

\section{Results} \label{Results}

In this section simulations of the agent based model described in section \ref{Method} are run in MATLAB 2020a. The dynamics of agent replication and evolutionary optimization of the phenotype are explored through these simulations. A set of standard conditions listed in Table \ref{standard cons} has been used in simulations throughout the results, while only varying one or two parameters per simulation.

\begin{table}[h!]
\centering
\begin{tabular}{|c|c|c|}
\hline
\multicolumn{3}{|c|}{\textbf{General parameters:}} \\
\hline
\textbf{Description} & \textbf{Symbol} & \textbf{Value} \\
\hline
Population size & $N$ & $1000$ \\
\hline
Initial mean fitness & $\Bar{f}(t=0)$ & $0.01$ \\
\hline
Initial mean curiosity & $\Bar{c}(t=0)$ & $0.1$ \\
\hline
Error rate & $\alpha$ & $0.01$ \\
\hline
Replication threshold (for all agent types) & $R_{replication}$ & $10$ \\
\hline
Learning cost & $R_{learning}$ & $0.1$ \\
\hline
Gene length (both curiosity gene and metabolic gene) & $L$ & $10$ \\
\hline
Maximum curiosity & $C$ & $1$ \\ 
\hline\hline
\multicolumn{3}{|c|}{\textbf{Additional parameters in time-dependent environment:}} \\
\hline
\textbf{Description} & \textbf{Symbol} & \textbf{Value} \\
\hline
Environment mutation rate & $\alpha_E$ & $0.5$ \\
\hline
Environment mutation period & $\tau_E$ & $100$ \\
\hline
\end{tabular}
\caption{Standard conditions for simulations}
\title{Standard conditions}
\label{standard cons}
\end{table}

\subsection{Mono agent type population fitness in constant environments} \label{section_f_dyn_E}
In this section the temporal dynamics of the mean fitness of pure populations in a constant environment are tested using simulations for all three agent types. First the general dynamics at standard conditions (Table \ref{standard cons}) are shown for the three agent types, whereafter the dynamics of the agent types are compared in simulations with chosen parameter variations.

\begin{figure}[H]
\centering
\begin{subfigure}[b]{1\textwidth}
    \advance\leftskip-1.25cm
    \includegraphics[scale=0.44]{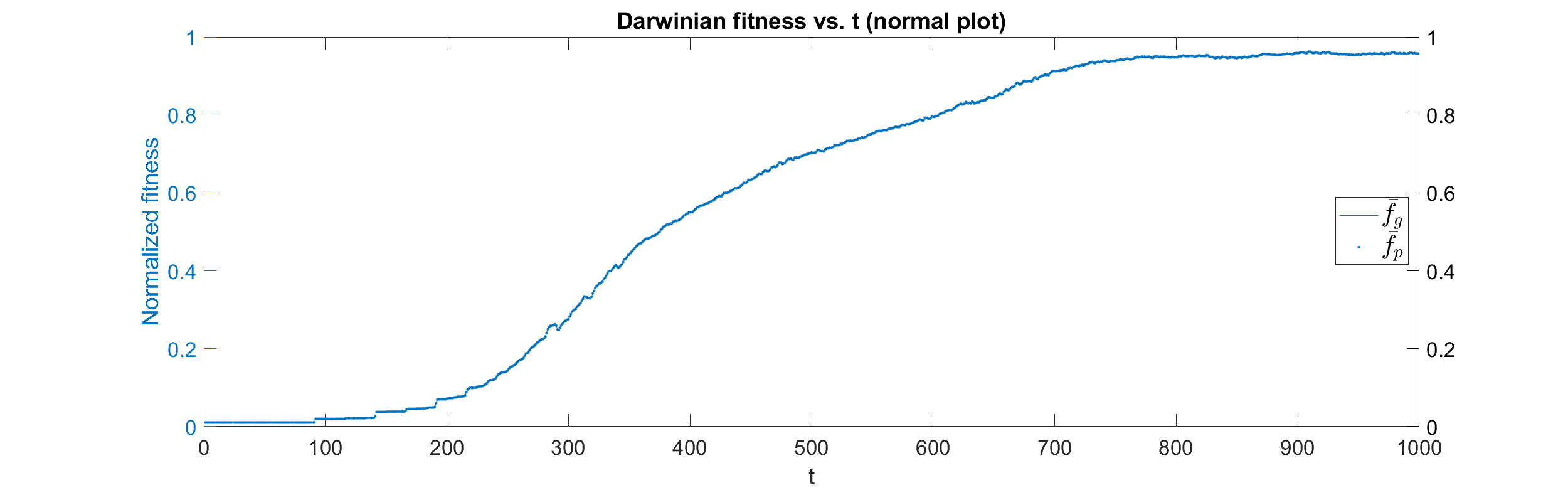}
\end{subfigure}
\hfill
\begin{subfigure}[b]{0.4\textwidth}
    \includegraphics[scale=0.40]{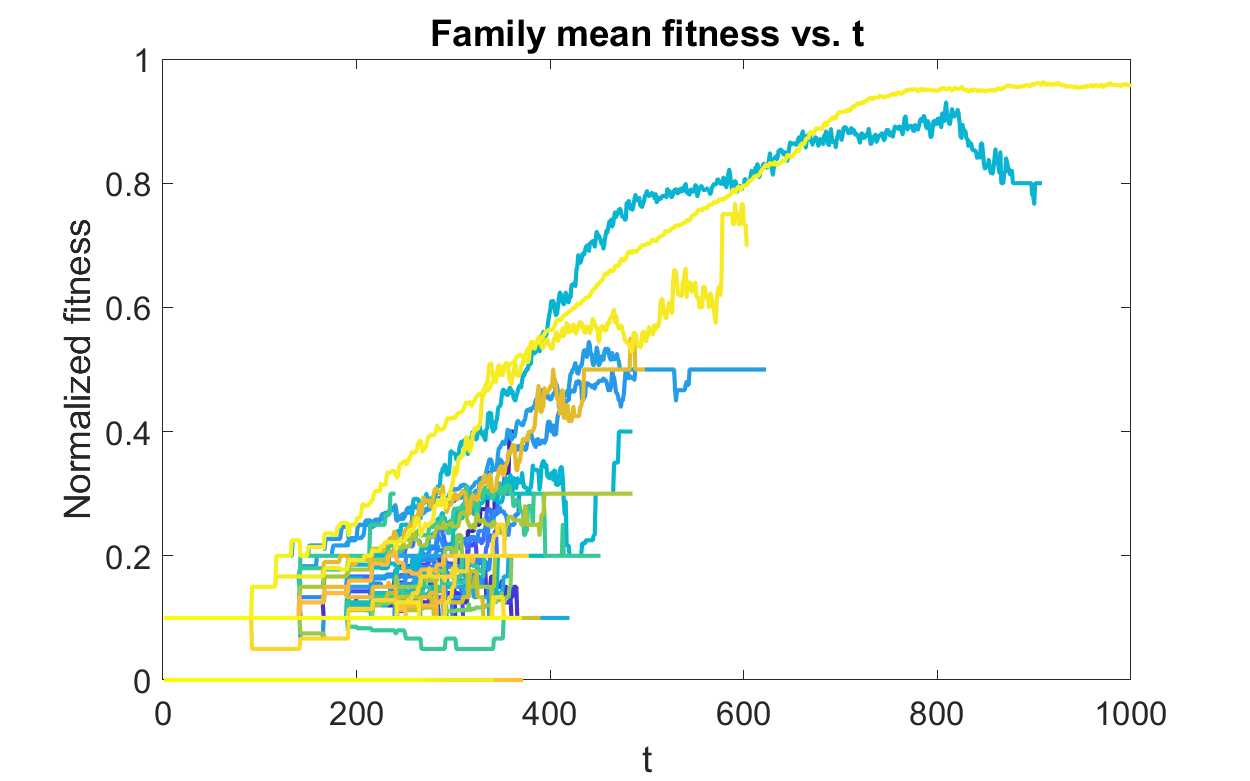}
\end{subfigure}
\hfill
\begin{subfigure}[b]{0.4\textwidth}
    \advance\leftskip-1.95cm
    \includegraphics[scale=0.40]{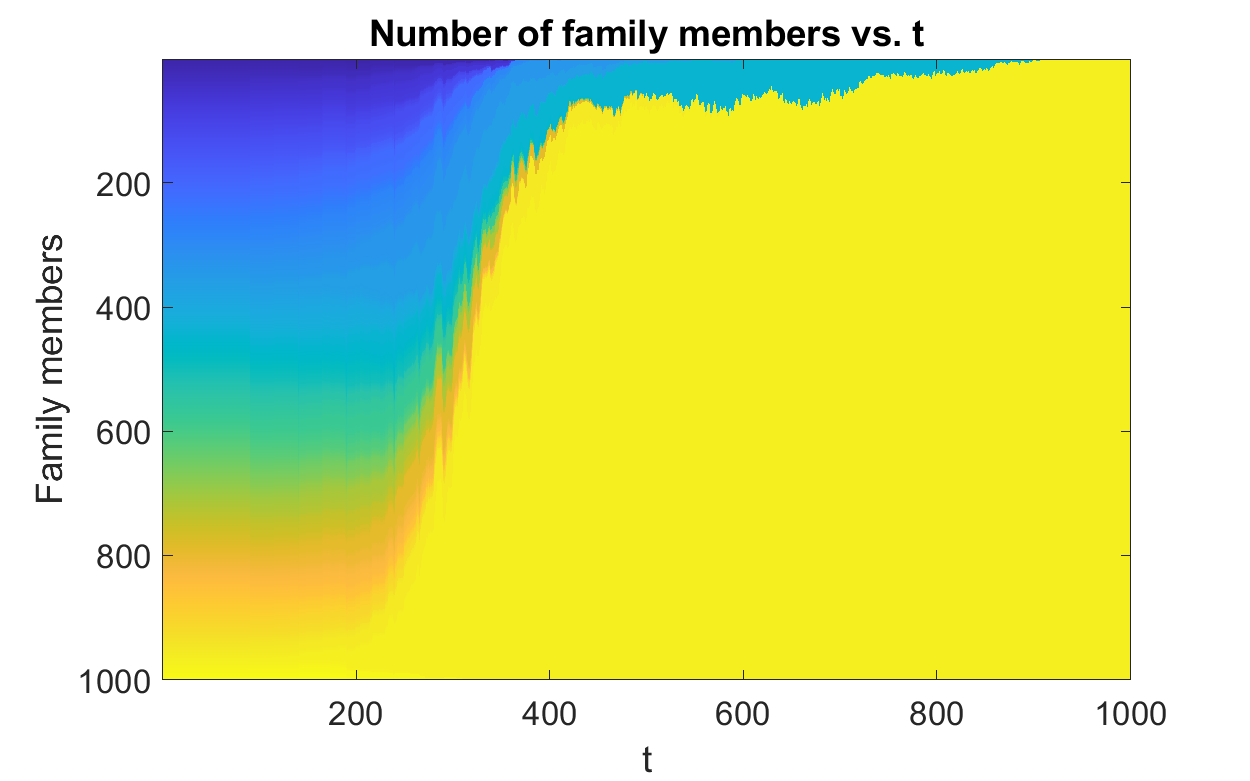}
\end{subfigure}
\caption{\textbf{Upper panel:} The temporal dynamics of population mean fitness of a pure Darwinian population plotted against $t$. \textbf{Lower left panel:} Family mean fitness of agent families in the pure Darwinian population vs. $t$. The individual agents in the initial population are ascribed a number from 1 to N - called their family identifier - with a corresponding color in the used colormap (blue $\rightarrow$ green $\rightarrow$ yellow). This identifier is inherited by their daughter agents. \textbf{Lower right panel:} The number of family members in each family vs. $t$. It should be noted that graphs may be hidden by other graphs at small $t$, as families mean fitness graphs are degenerate for Darwinian populations because of the small variance in phenotypic fitness (Figure \ref{Var f_p}).}
\label{D f dyn}
\end{figure}

 For Darwinian agents the mean phenotypic fitness graph and the mean genotypic fitness graph overlays, as the genotype maps directly to the phenotype, as explained in Section \ref{section_Darwinian_agents}. Initially, the optimization is very slow as the initial mean fitness of $0.01$ means only $\approx$10\% of the initial population have a non-zero fitness. At low fitness, the optimization of Darwinian agents are slower than learning-based algorithms, since the rate of optimization is dependent on replication rate, while  At low mean fitness the optimization is slow since the agents must first collect enough resources to replicate. At high mean fitness, the optimization may be slowed by coincidental replacement of agents with higher than mean fitness by agents with lesser than mean fitness due to the high flux of new agents. For the used simulation, the family with the identifier 970 (bright yellow) becomes the dominating family due to its higher rate of replication caused by its high fitness.

\begin{figure}[H]
\centering
\begin{subfigure}[b]{1\textwidth}
    \advance\leftskip-1.25cm
    \includegraphics[scale=0.44]{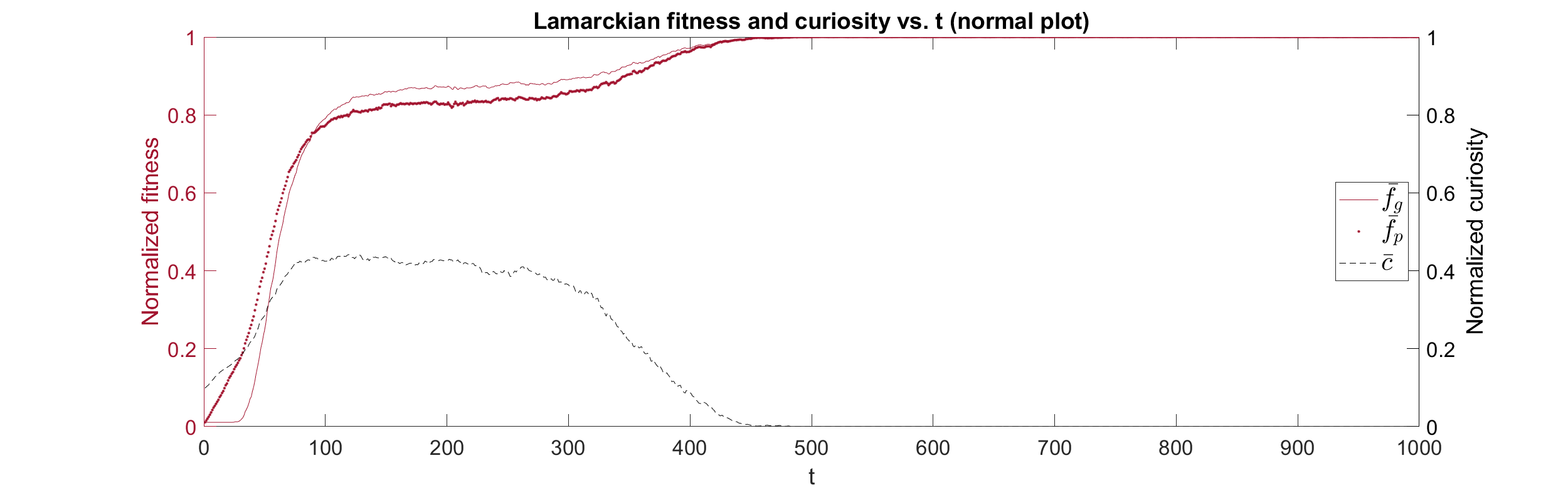}
\end{subfigure}
\hfill
\begin{subfigure}[b]{0.4\textwidth}
    \includegraphics[scale=0.40]{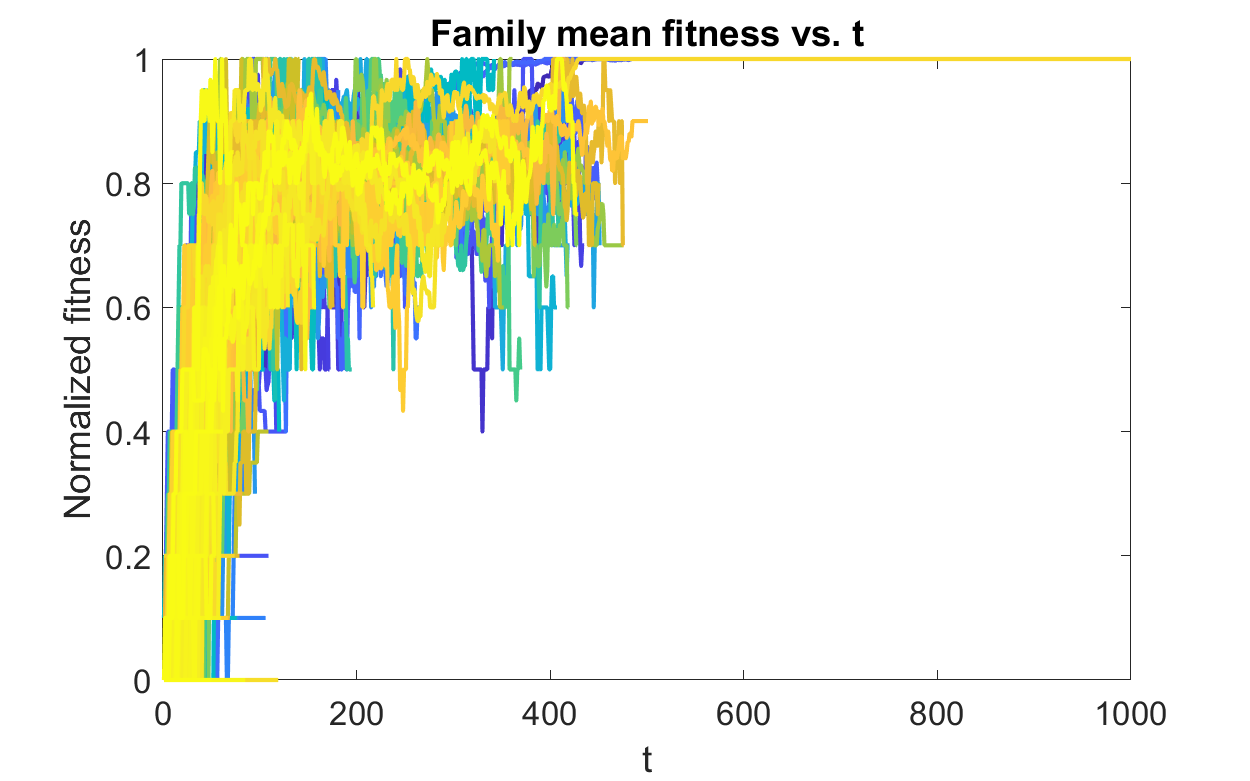}
\end{subfigure}
\hfill
\begin{subfigure}[b]{0.4\textwidth}
    \advance\leftskip-1.95cm
    \includegraphics[scale=0.40]{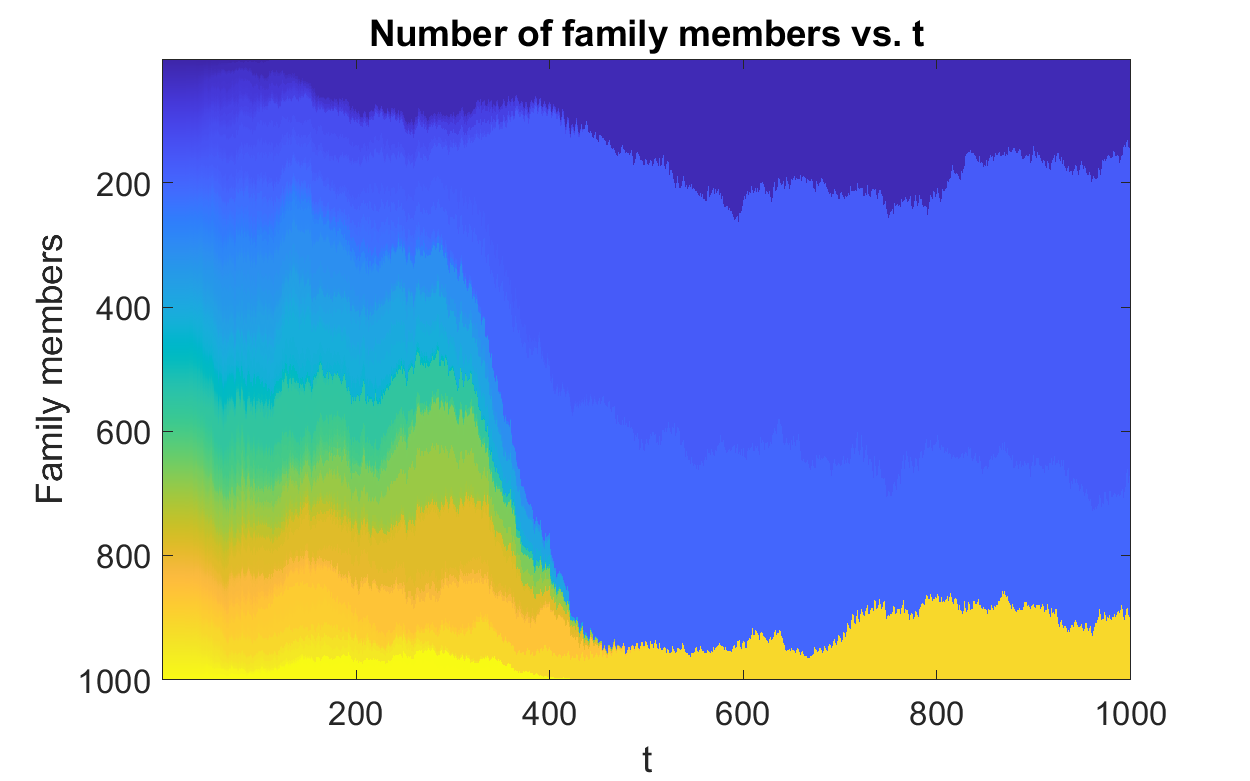}
\end{subfigure}

\caption{\textbf{Upper panel:} The temporal dynamics of population mean fitness of a pure Lamarckian population plotted against $t$. \textbf{Lower left panel:} Family mean fitness of agent families in the pure Lamarckian population vs. $t$. \textbf{Lower right panel:} The number of family members in each family vs. $t$.}
\label{L f dyn}
\end{figure}

At standard conditions the mean fitness of the Lamarckian population increases significantly quicker than the other agent types. Learning events efficiently increase the phenotypic fitness (dotted red line) at small $t$ as the likelihood of switches increasing the fitness is large. The genotypic fitness (thin red line) starts increasing as the agents begin to replicate at $t\approx 25$. A selection pressure for high curiosity is initially present, resulting in a high mean curiosity (dashed black line). As the population mean fitness increases, learning events become less efficient at increasing phenotypic fitness, as agents become more likely to decrease their fitness from switches. After a critical point at $t\approx 100$ we have $\Bar{f}_p=\Bar{f}_p$, and the dynamics change as the high population mean curiosity instead result in $\Bar{f}_p(t>100)<\Bar{f}_g(t>100)$, and a selection pressure for lower curiosity is generated. At $t>300$ the mean curiosity decreases as agents with low curiosity and high fitness emerge, and the system finally reaches a constant final state of a population consisting purely of agents with a single high fitness phenotype and zero curiosity. For a genome length of $L=10$ - which is the used standard condition - this is likely to be the state of maximum fitness, but for larger genomes the final agents are unlikely to have exactly maximum fitness.

Comparing the family dynamics of the Lamarckian population with the family dynamics of the Darwinian population shown on Figure \ref{D f dyn} middle panel reveals a higher variation of phenotypic fitness within each family as learning events increase the plasticity of the phenotype (also illustrated in Figure \ref{Var f_p}). At $t\approx 500$, all agents in the population has maximal fitness. As seen on \autoref{L f dyn} Lower Panel, simulations may result in multiple competing families with maximal fitness.

\begin{figure}[H]

\centering
\begin{subfigure}[b]{1\textwidth}
    \advance\leftskip-1.25cm
    \includegraphics[scale=0.44]{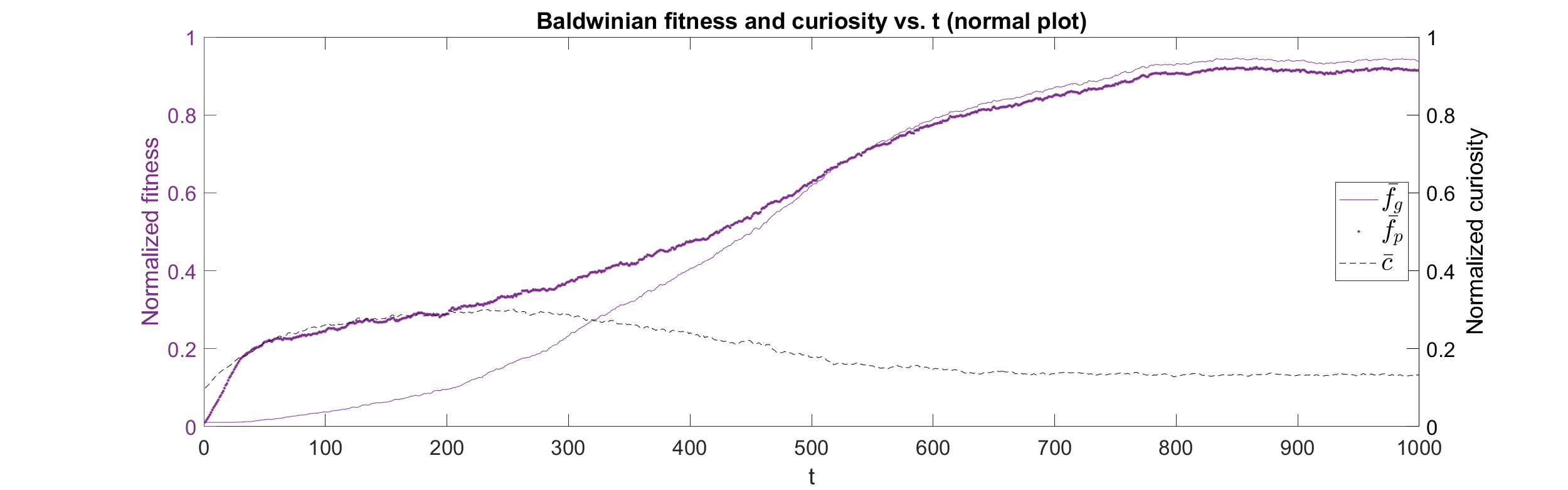}
\end{subfigure}
\hfill
\begin{subfigure}[b]{0.4\textwidth}
    \includegraphics[scale=0.40]{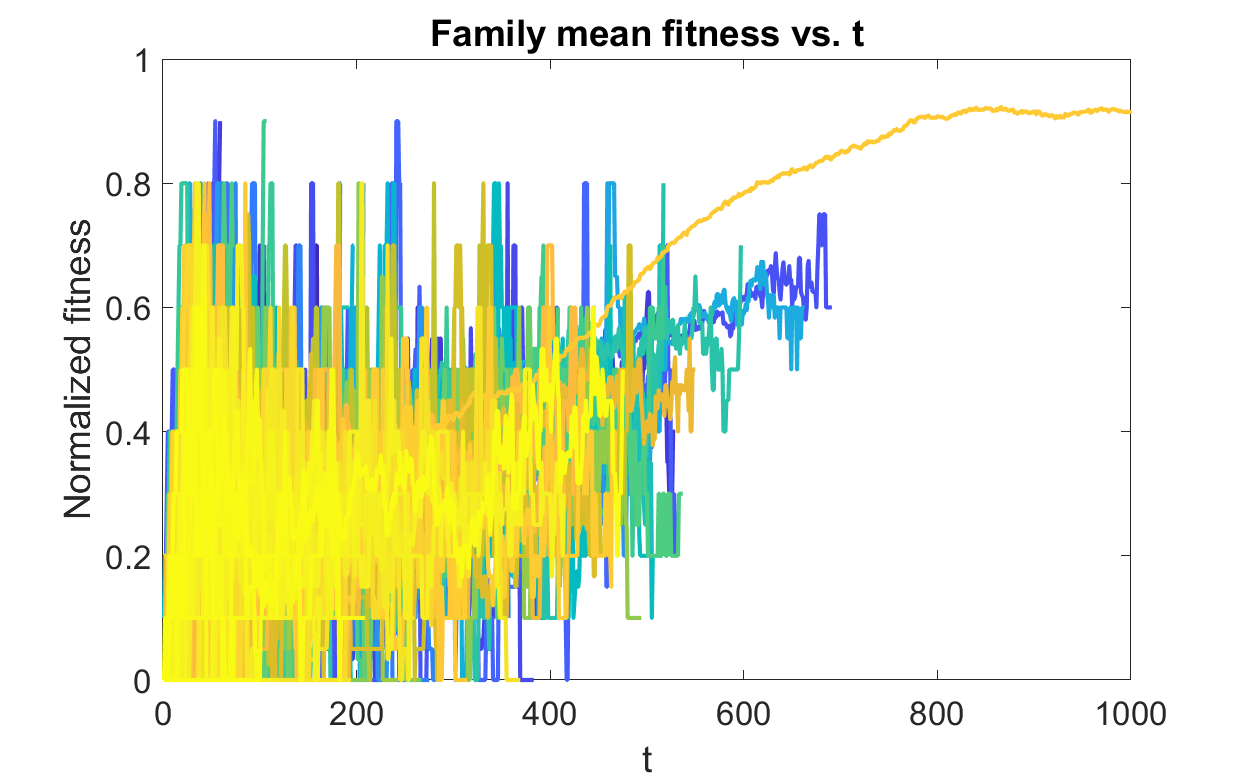}
\end{subfigure}
\hfill
\begin{subfigure}[b]{0.4\textwidth}
    \advance\leftskip-1.95cm
    \includegraphics[scale=0.40]{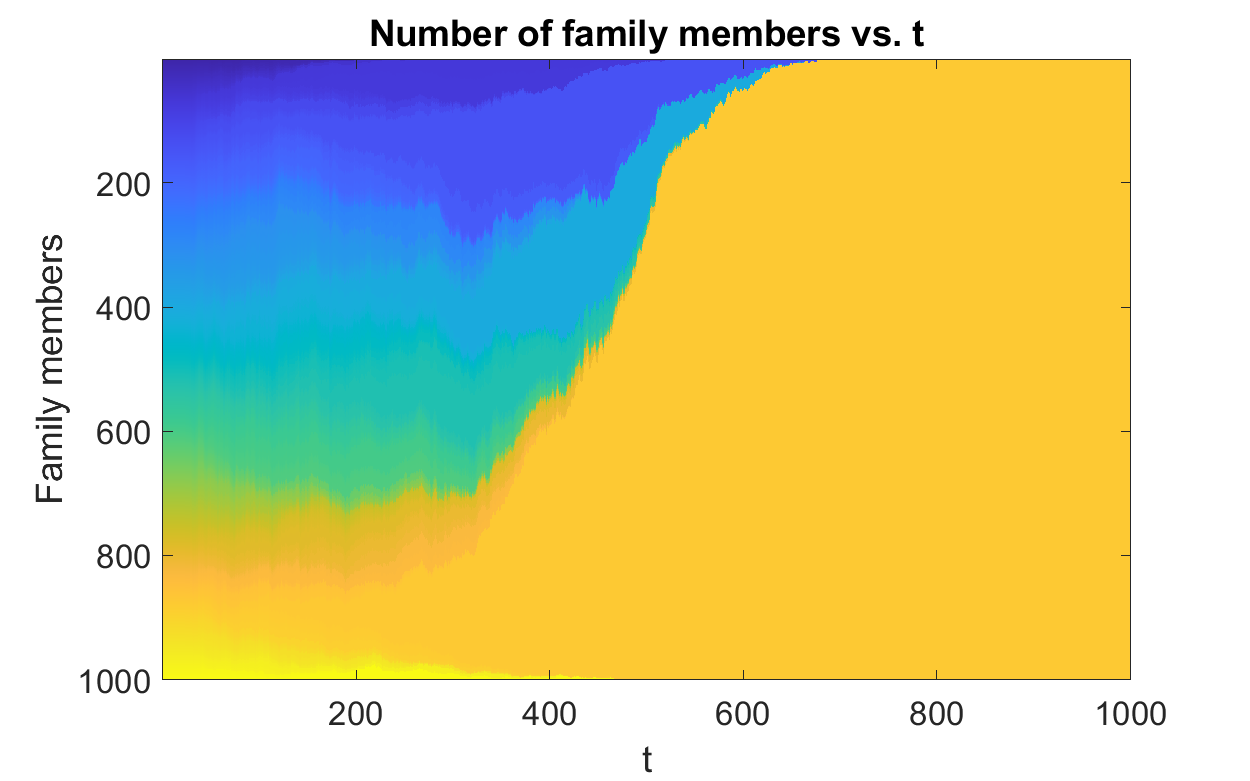}
\end{subfigure}
\caption{\textbf{Upper panel:} The temporal dynamics of population mean fitness of a pure Baldwinian population plotted against $t$. \textbf{Lower left panel:} Family mean fitness of agent families in the pure Baldwinian population vs. $t$. \textbf{Lower right panel:} The number of family members in each family vs. $t$.}
\label{B f dyn}
\end{figure}

The Baldwinian population initially quickly increases its mean phenotypic fitness via learning like the Lamarckian population. However, this only boosts the fitness of Baldwinian agents during their lifetime, and replicated agents do not benefit. The genotypic fitness can only increase through random mutations during replication. Since Baldwinian agents spend resources on learning, the replication rates of Baldwinian agents are lower than for Darwinian agents, and the rise in mean genotypic fitness is therefore slower than for the Darwinian population shown on Figure \ref{D f dyn}. Learning becomes less efficient as phenotypic fitness increases, and like the Lamarckian population, the Baldwinian population reaches a critical point at $t\approx 425$ whereafter $\Bar{f}_p(t>100)<\Bar{f}_g(t>100)$. Unlike the Lamarckian population, the Baldwinian population is not able to find a final state consisting of agents with high fitness and zero curiosity, as the random mutations introduce noise into both the curiosity and metabolic gene pools.

Similar to the Lamarckian population on Figure \ref{L f dyn}, the Baldwinian population have a high variation of phenotypic fitness within each family due to the phenotypic plasticity generated by learning.


\begin{figure}[H]
\centering
\advance\leftskip-2cm
\includegraphics[scale=0.48]{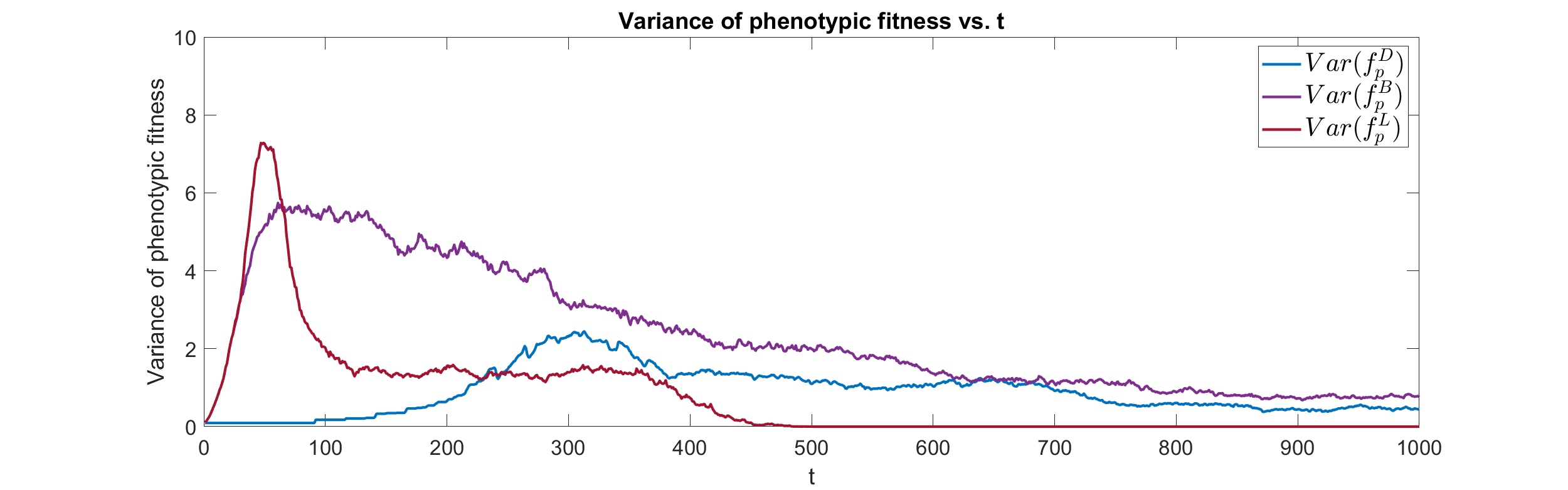}
\caption{The variance of phenotypic fitness in the populations from the simulations shown on Figures \ref{D f dyn}-\ref{B f dyn}. The Lamarckian population (red) shows a very high initial increase in phenotypic variance followed by a correspondingly fast decrease, as the population has reached a mean fitness of $\sim 0.8$. At $t\approx 490$, the Lamarckian phenotypic variance becomes zero, as all agents in the population have optimal fitness and zero curiosity. The Darwinian population (blue) has the lowest maximal phenotypic variance; and this maximum also happens later in the dynamics than for the Lamarckian and Baldwinian populations. The Baldwinian population (purple) shows higher phenotypic variance compared to the Darwinian population at all times, as the Baldwinian agent type was designed to do.}
\label{Var f_p}
\end{figure}

\subsubsection{Comparison of mono agent type populations: constant environment} \label{sec:comparison}

\begin{figure}[H]
\centering
\advance\leftskip-1.25cm
\includegraphics[scale=0.44]{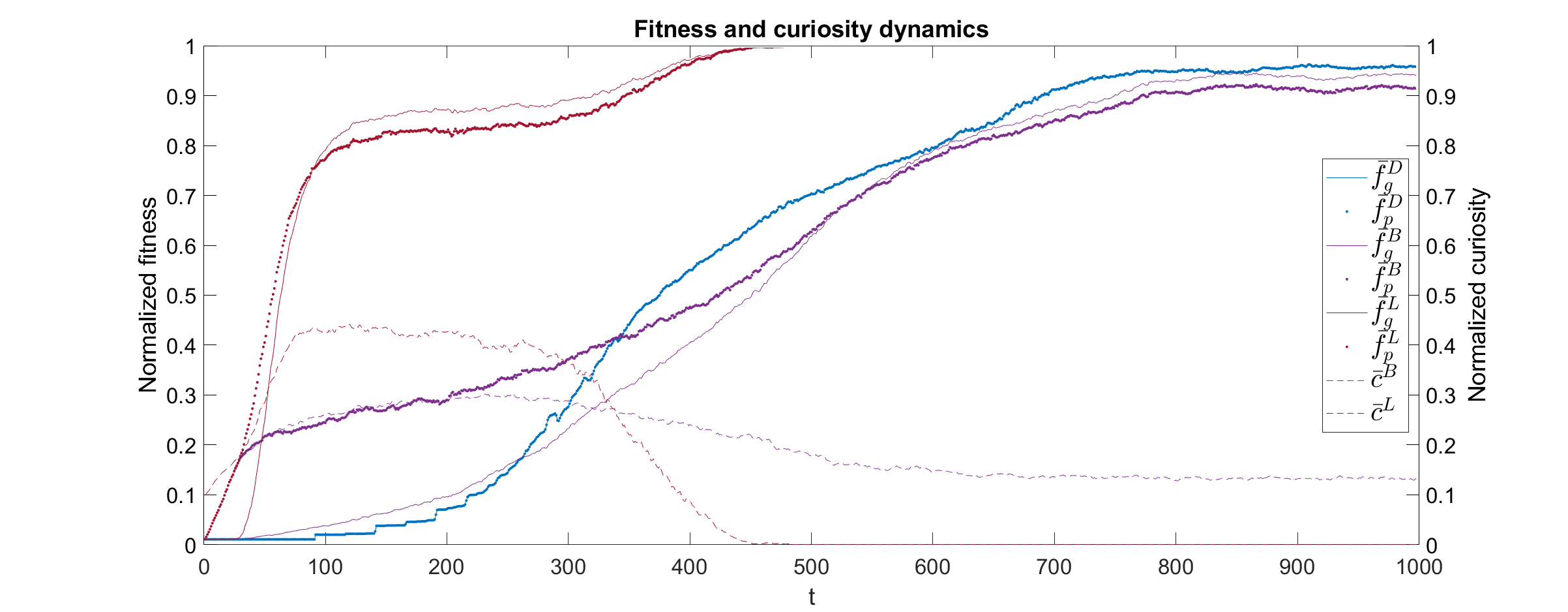}
\caption{Standard conditions. Darwinian computation time = 14 seconds, Baldwinian computation time = 21 seconds, Lamarckian computation time = 23 seconds.}
\label{N=1000}
\end{figure}

\begin{figure}[H]
\centering
\advance\leftskip-1.25cm
\includegraphics[scale=0.44]{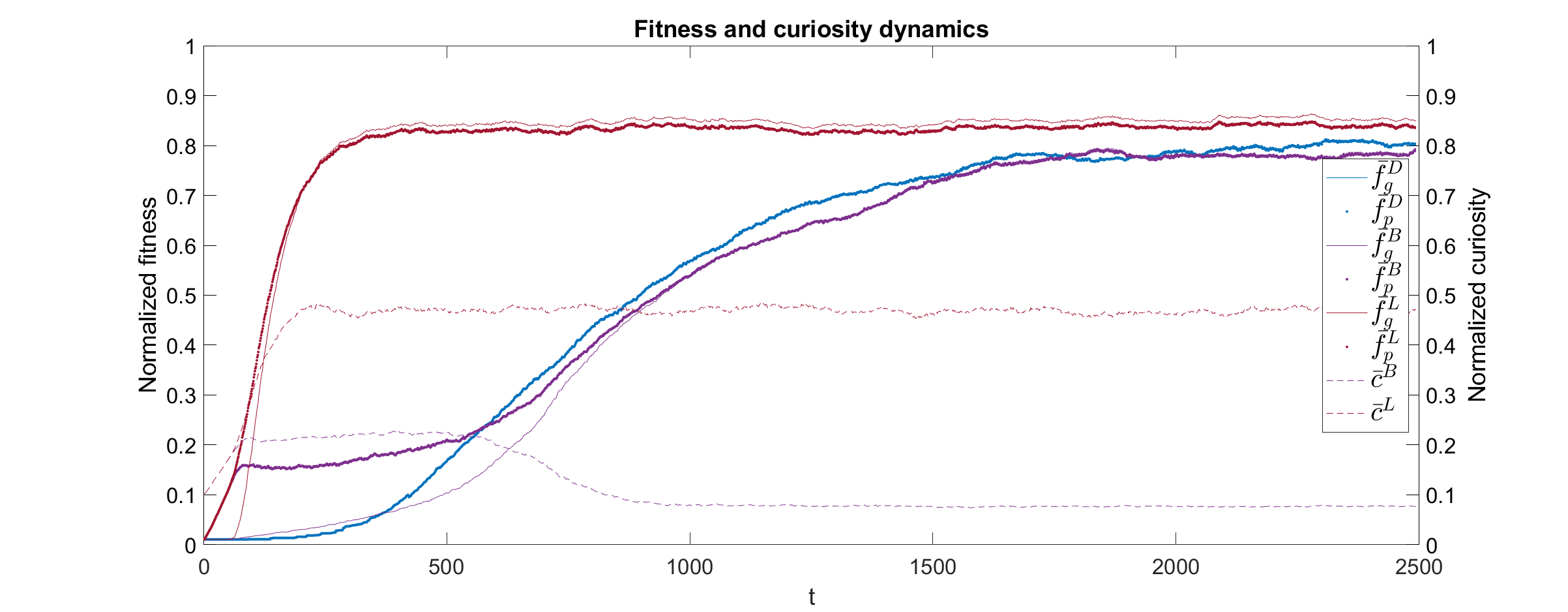}
\caption{Standard conditions with $L=30$. At larger gene lengths, the Lamarckian population is less likely to enter the part of the dynamics where the curiosity is down-regulated within the duration of the simulation. Darwinian computation time = 39 seconds, Baldwinian computation time = 57 seconds, Lamarckian computation time = 86 seconds.}
\label{N=1000 L=30}
\end{figure}

\begin{figure}[H]
\centering
\advance\leftskip-1.25cm
\includegraphics[scale=0.44]{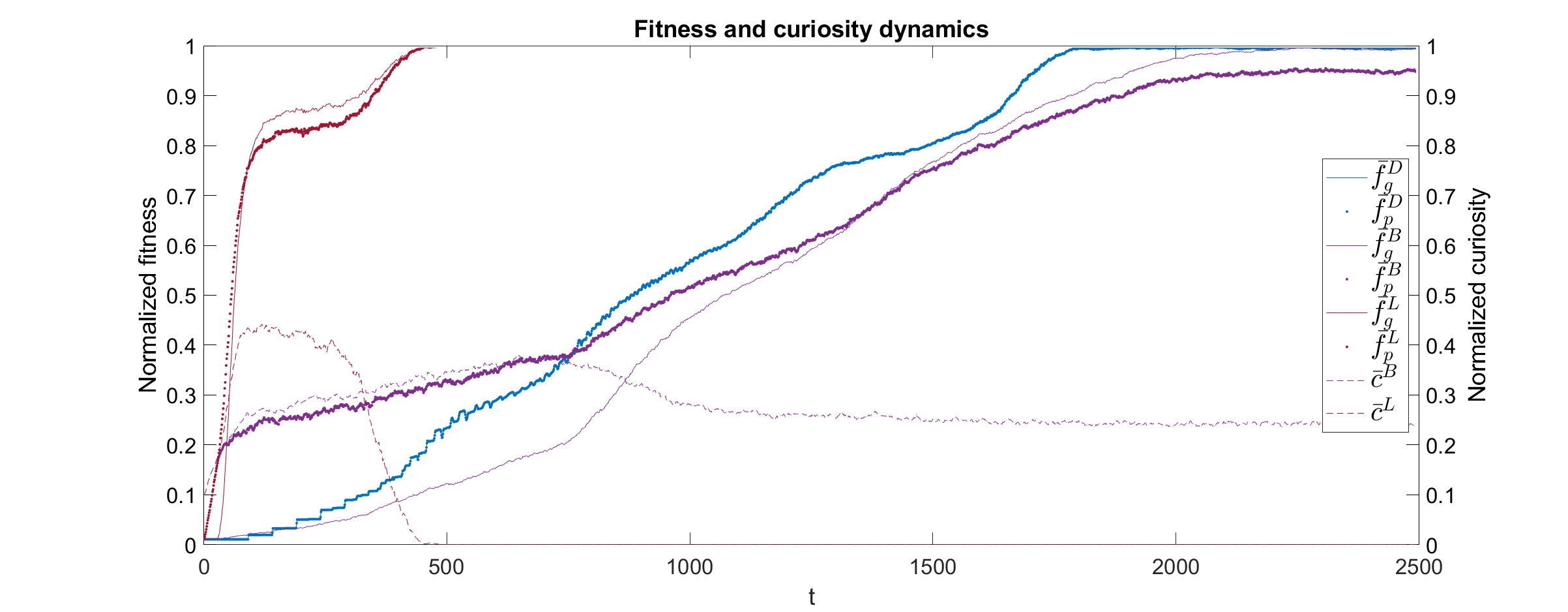}
\caption{Standard conditions with $\alpha=0.001$. With a smaller error rate, the Darwinian and Baldwinian optimization is slower, but the Darwinian population is able to reach states of higher fitness likely due to the lesser noise introduced in the population. The Baldwinian population is inhibited from reaching this state of high fitness due to the noise introduced by learning events, though given enough simulation time, the population may be able to down-regulate curiosity. Darwinian computation time = 35 seconds, Baldwinian computation time = 53 seconds, Lamarckian computation time = 55 seconds.}
\label{N=1000 alpha=0.001}
\end{figure}

\begin{figure}[H]
\centering
\advance\leftskip-1.25cm
\includegraphics[scale=0.44]{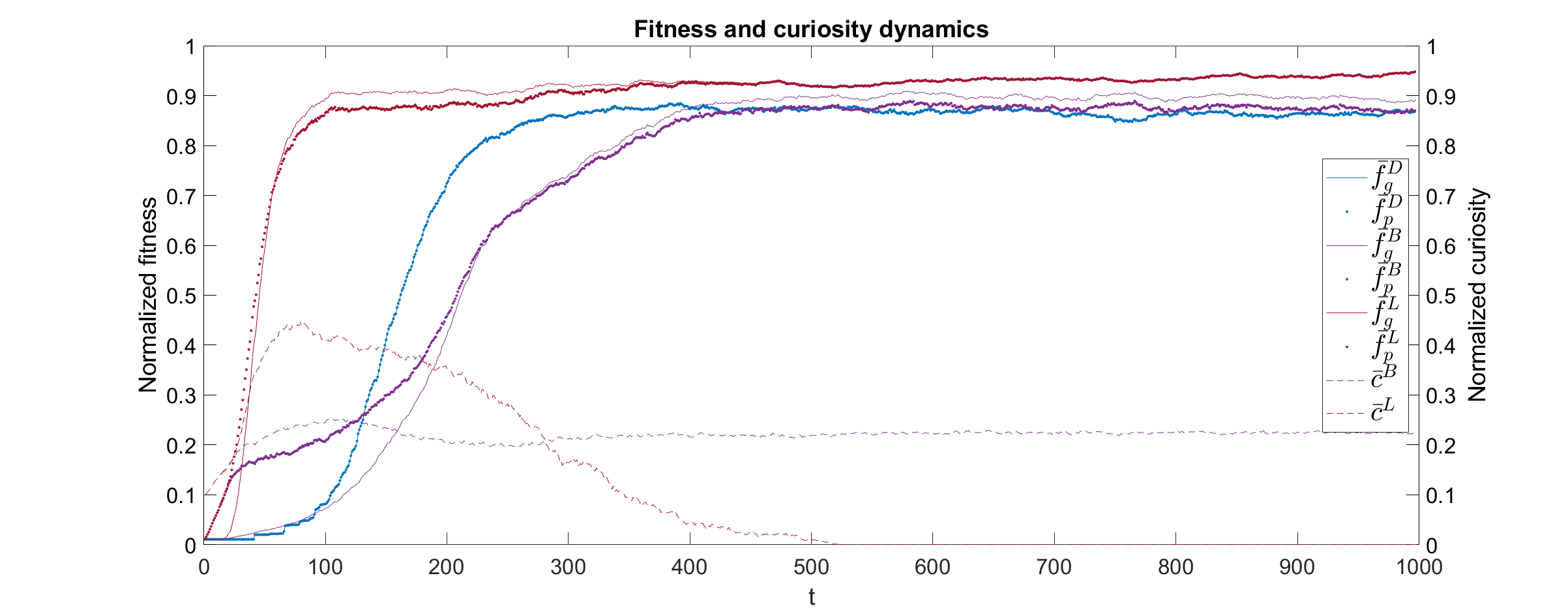}
\caption{Standard conditions with $R_{replication}=5$. At a smaller replication threshold, the Darwinian algorithm is able to search through more states. The optimization is therefore faster.  The non-constant fitness of the Lamarckian population after reaching zero mean curiosity is due to the presence of several competing families with zero curiosity at $t>520$. Darwinian computation time = 20 seconds, Baldwinian computation time = 26 seconds, Lamarckian computation time = 30 seconds.}
\label{N=1000 R=5}
\end{figure}

\begin{figure}[H]
\centering
\advance\leftskip-1.25cm
\includegraphics[scale=0.44]{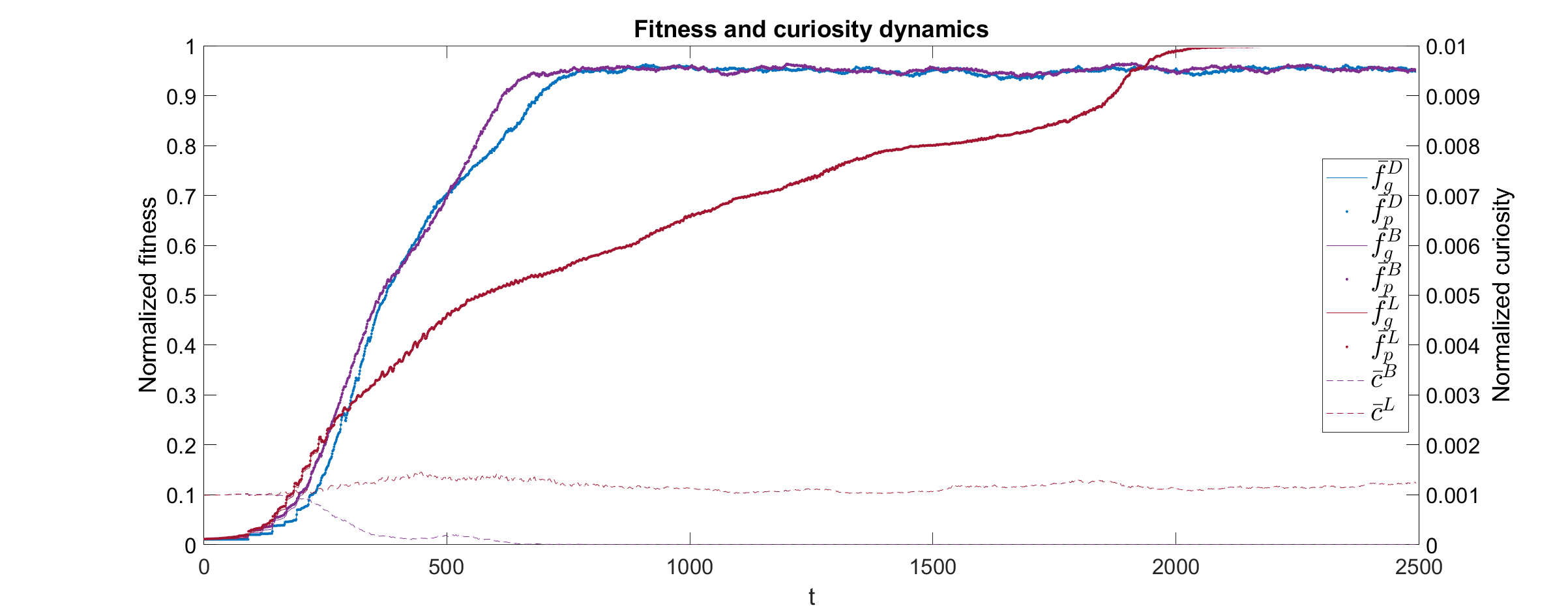}
\caption{Standard conditions with $C=0.01$. At low maximum curiosity, the optimization of the Lamarckian population is slower than the Darwinian and Baldwinian. The Baldwinian dynamics are nearly identical to the Darwinian dynamics, as the Baldwinian agent type coincides with the Darwinian at zero curiosity, as explained in Section \ref{sec: B method}. Darwinian computation time = 40 seconds, Baldwinian computation time = 55 seconds, Lamarckian computation time = 55 seconds.}
\label{N=1000 C=0.01}
\end{figure}

\begin{figure}[H]
\centering
\advance\leftskip-1.25cm
\includegraphics[scale=0.44]{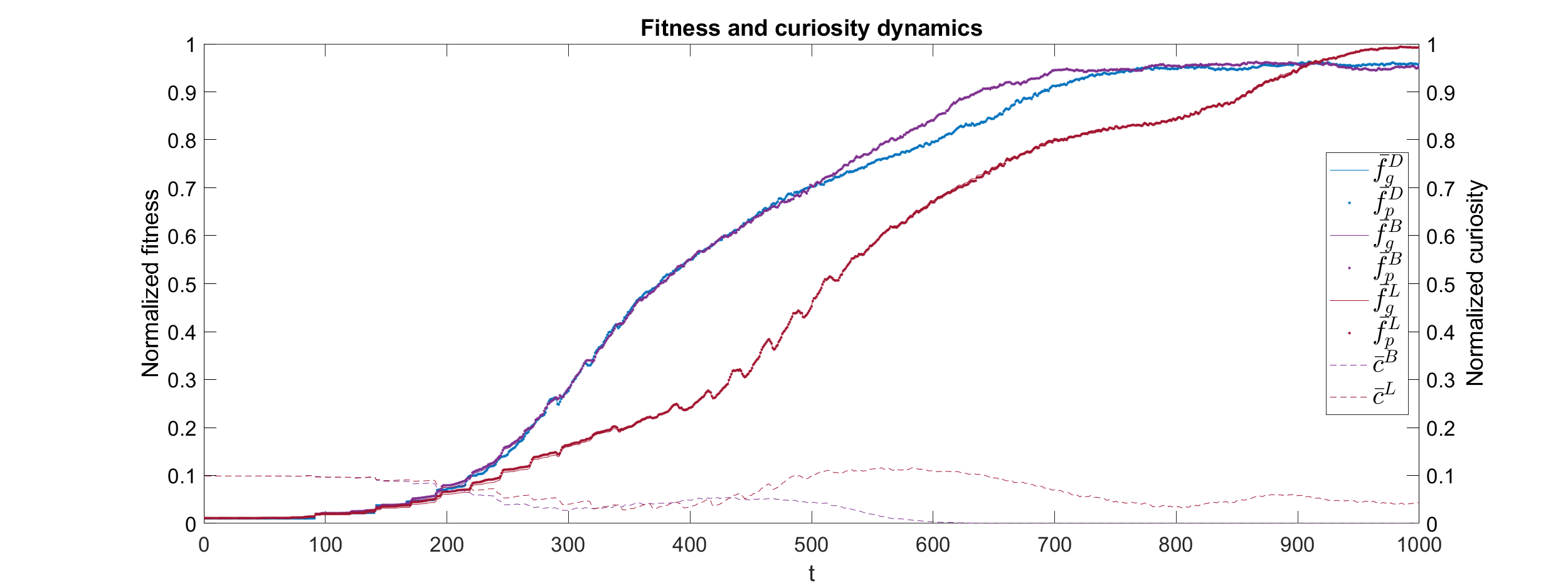}
\caption{Standard conditions with $R_{learning}=9$. At learning cost comparable to the replication cost, the optimization of the Lamarckian population is slower than the Darwinian and Baldwinian. A selection pressure for lowering curiosity emerges in the Baldwinian population. Darwinian computation time = 15 seconds, Baldwinian computation time = 21 seconds, Lamarckian computation time = 21 seconds.}
\label{N=1000 R_L=9}
\end{figure}

In this section, the fitness and curiosity dynamics of pure populations of all three agent types are compared while changing a single parameter from its standard value and keeping all other parameters as defined in Table \ref{standard cons}. Here, pure populations mean that three separate simulations with either 1000 Darwinian, Lamarckian or Baldwinian agents have been run with a given set of parameters, and the results of these three simulations have been plotted in a single figure for direct comparison of phenotypic and genotypic fitness optimization dynamics as well as the mean curiosity dynamics. In the Figures \ref{N=1000}-\ref{N=1000 R_L=9} the parameters $L$, $\alpha$, $R_{replication}$, $C$, and $R_{learning}$ are varied consecutively. The simulation runtimes are also shown to gauge and compare the computational cost of the simulations. Depending on the simulation, the total number of $t$-steps ($T$) is either $T=1000$ or $T=2500$, which should be regarded while comparing figures and runtimes.

To summarize, at the chosen standard conditions the Lamarckian agent type optimizes fitness better than both the Darwinian and Baldwinian agent types. For larger gene length ($L=30$) all agent types optimize slower and reach a lower final fitness within the allotted simulation time. Furthermore, the Lamarckian agent type is less likely to find a state of near-maximal fitness and zero curiosity, and the dynamics therefore does show the characteristic second rise in fitness. At lower mutation rate ($\alpha=0.001$) yields slower optimization for the Darwinian and Baldwinian agent types, but also a higher optimal fitness due to the lesser noise introduced by mutations. If the mutation rate instead increases, the Darwinian and Baldwinian agent type optimize faster, but reach a lower optimal fitness. At standard conditions, we do not observe a crossover - where the Darwinian and Baldwinian agent types beat the Lamarckian agent type - by only varying the mutation rate $\alpha$. Lower replication cost benefits the Darwinian and Baldwinian agent types, as their optimization rates scale with replication rate, while the Lamarckian optimization rate stays constant. At standard conditions, we do not observe a crossover by only varying the replication cost. At low maximum curiosity ($C=0.01$) the optimization rate of the Lamarckian agent type is lowered drastically, and a crossover in the dynamics is observed. Finally, increasing the learning cost slows Lamarckian optimization. However, a crossover in the dynamics is only observed at very high learning costs, i.e., when the learning cost is comparable to the replication cost ($R_{learning}=9\approx R_{replication} = 10$).

\subsection{Mono agent type population fitness in temporally varying environment}\label{section_f_dyn_E(t)}

Now the fitness dynamics for pure populations of all three agent types are investigated in a temporally dependent environment as defined in Section \ref{sec: E(t)}. The relative effectiveness of the agent types is measured by comparing the time-average of the mean fitness during steady state of the dynamics; steady state being defined as $t>3000$.


\begin{figure}[H]
\advance\leftskip-3cm
\includegraphics[scale=0.46]{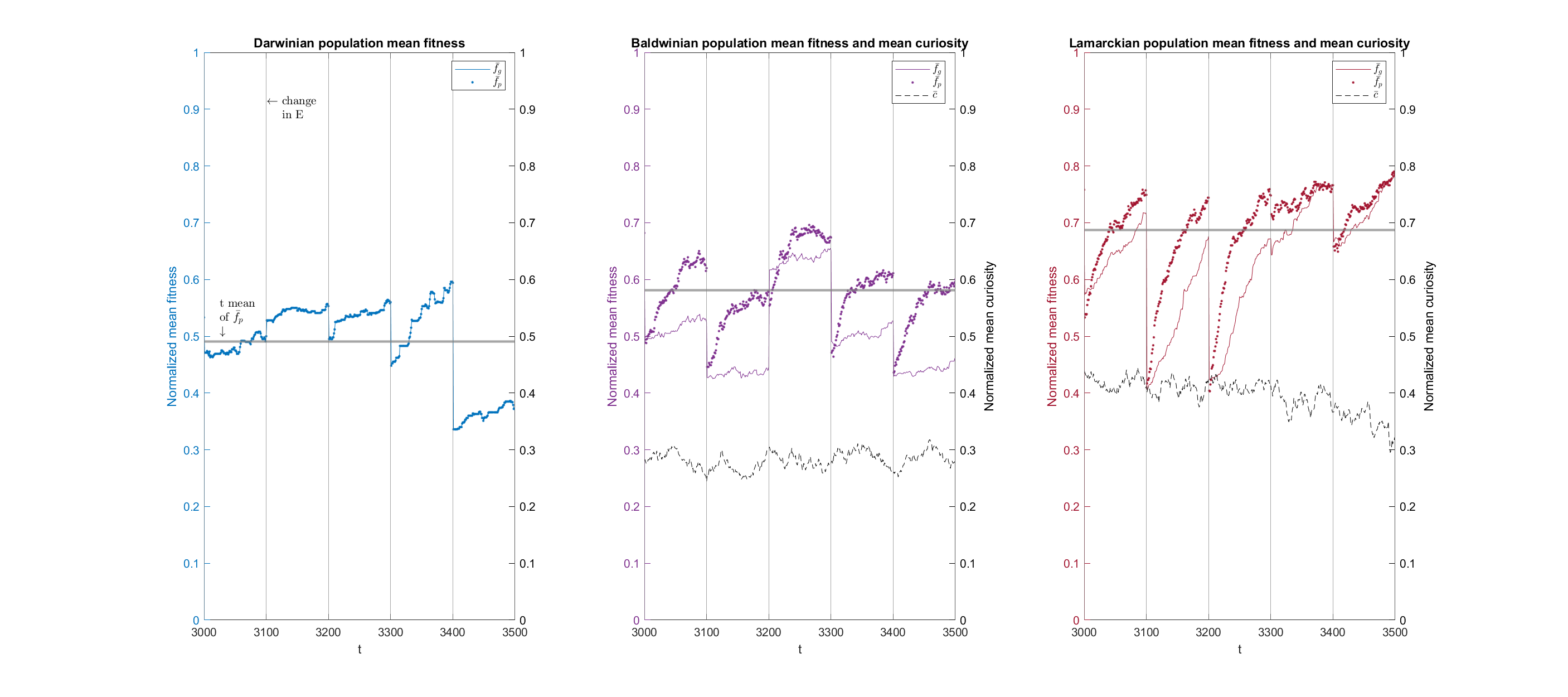}
\caption{A comparison of the fitness dynamics of pure populations in a temporal environment defined in Section \ref{sec: E(t)}. The parameters are defined such that the environment is scrambled ($\alpha_E=0.5$) every 100 t-steps ($\tau_E=100$). Parameter values used: Standard conditions except $R_{replication}=50$ and $N=100$.}
\label{temporalE}
\end{figure}

For the parameters used in the simulation on Figure \ref{temporalE}, the Lamarckian agent type has the highest time-averaged mean phenotypic fitness (horizontal gray line), followed by the Baldwinian agent type, and finally the Darwinian agent type has the lowest time-averaged mean fitness. The simulations show that when the replication rate is on the same time scale as the mutations of the environment, the Darwinian method of increasing fitness through replication with mutation is not a viable strategy for evolutionary optimization. Lamarckian and Baldwinian populations are simply able to explore more states utilizing the curiosity-based learning algorithm before the environment mutates. 20 simulations with the same parameters as used for the simulation on Figure \ref{temporalE} but with varying rng-seeds show the general results for the fitness dynamics at steady state ($3000<t<6000$): Time-averaged mean phenotypic fitness $\Bar{f}_p^L=0.6701>\Bar{f}_p^B=0.6013>\Bar{f}_p^D=0.5166$; Time-averaged mean genotypic fitness $\Bar{f}_g^L=0.6022>\Bar{f}_g^D=0.5166>\Bar{f}_g^B=0.5057$. The highest mean genotypic fitness is reached by the Lamarckian population due to transmission of the phenotypic changes to the genotype of the new agent, i.e., inheritance of acquired traits, and the genotypic mean fitness of the Baldwinian and Darwinian populations are approximately equal, due to the lack thereof. This is illustrated on Figure \ref{temporalE}.

\subsection{Pairwise agent population competition}\label{Pairwise}

In this section, the total population is allowed to be a mix of different agent types, allowing for simulations of direct competition between agent types. For such simulations with a finite total population size, the population drifts towards homogeneity, meaning the total population will consist of a single agent type within a finite amount of $t$-steps. This final agent type is then considered to be winner.

\begin{figure}[H]
\centering
\advance\leftskip-1.25cm
\includegraphics[scale=0.44]{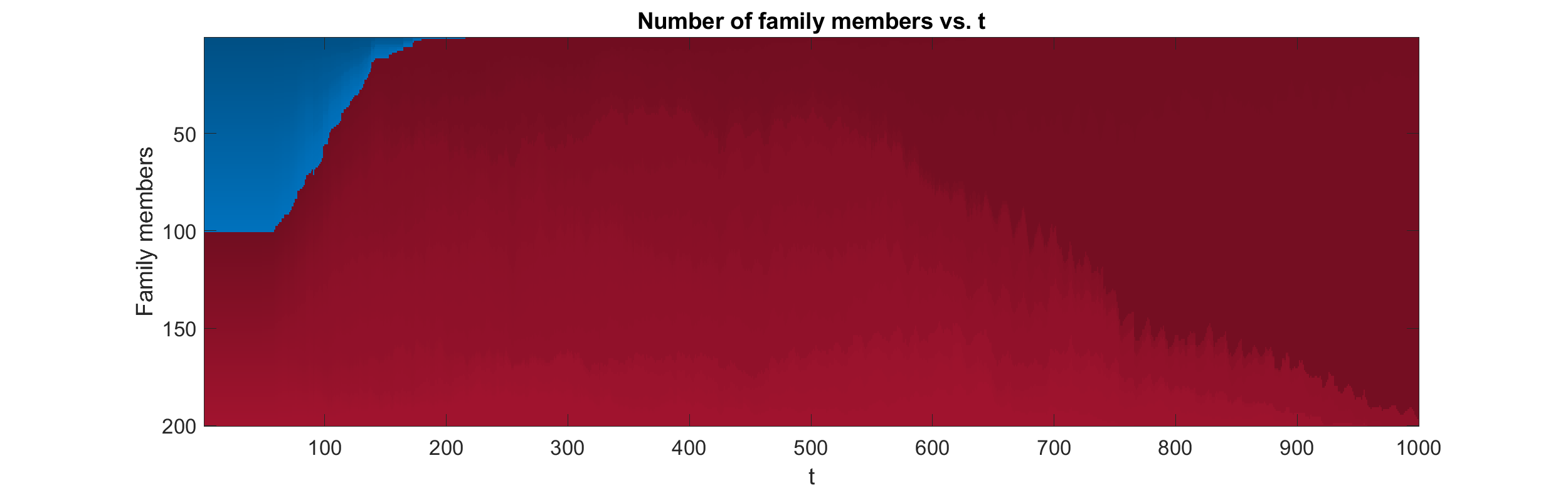}
\caption{Family dynamics of a mixed initial population consisting of 100 Darwinian agents (blue spectrum) and 100 Lamarckian agents (red spectrum). For the used parameters and initial conditions, the Lamarckian agent type fills the population at $t\approx 200$, and is therefore declared the winning agent type. Furthermore, at $t\approx 1000$, only one Lamarckian family dominates the population, as shown by the complete population being composed of a single shade of red.}
\label{comp_fam_dyn}
\end{figure}

For a non-analytic model as ours, a complete analysis of the outcomes of such competitions would be obtained by running simulations while varying every parameter and initial condition, making a high dimensional phase space, wherein each point has a single winning agent type. Furthermore, to make the results rigorously statistically significant, multiple simulations would have to be run for each point in the phase space, since the random number generation influences the final outcome; especially close to the interfaces between areas where one agent has a clear advantage. This is not possible in practise, and the results would be equally hard to illustrate.

Instead, we have chosen to limit our simulations by varying two parameters at a time, keeping all other parameters and initial conditions equal to the standard conditions specified in Table \ref{standard cons}, while only allowing two agent types to compete in each simulation with their initial populations being equal. For a constant environment the chosen parameters are the learning cost $R_{learning}$ and the relative replication cost $R_{ij}$ between the agent types $i$ and $j$, which is defined as $R_{ij}=R^i_{replication}/R^j_{replication}$. A simulation is run and the winning agent type is found for each sampled pair of parameter values corresponding to the pixels in the subsequent figures. The colors of the pixels in these figures correspond to the winning agent types using the established color convention.

\begin{figure}[H]
\begin{minipage}{0.32\textwidth}
\centering
\includegraphics[scale=0.38]{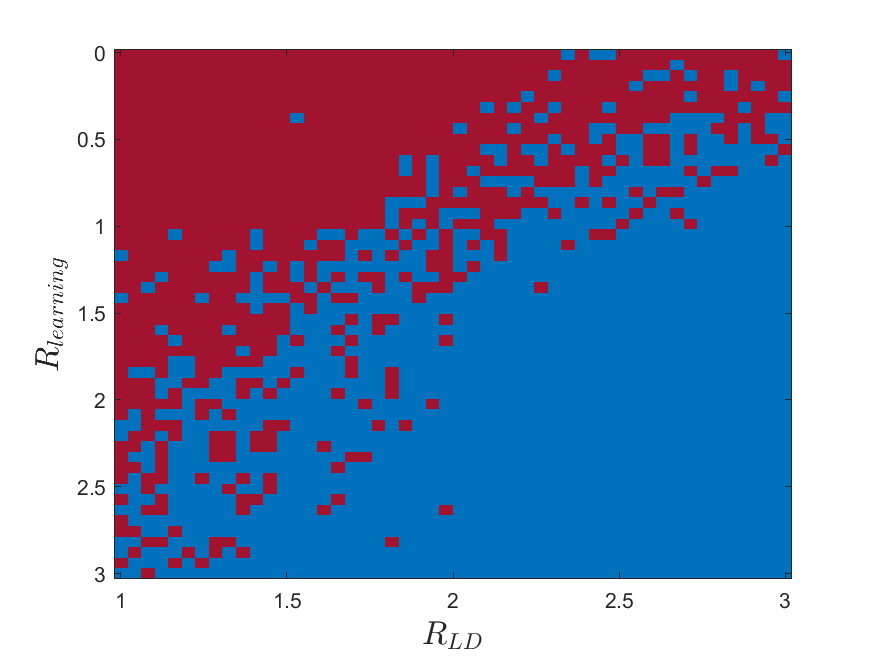}
\end{minipage}
\begin{minipage}{0.32\textwidth}
\centering
\includegraphics[scale=0.38]{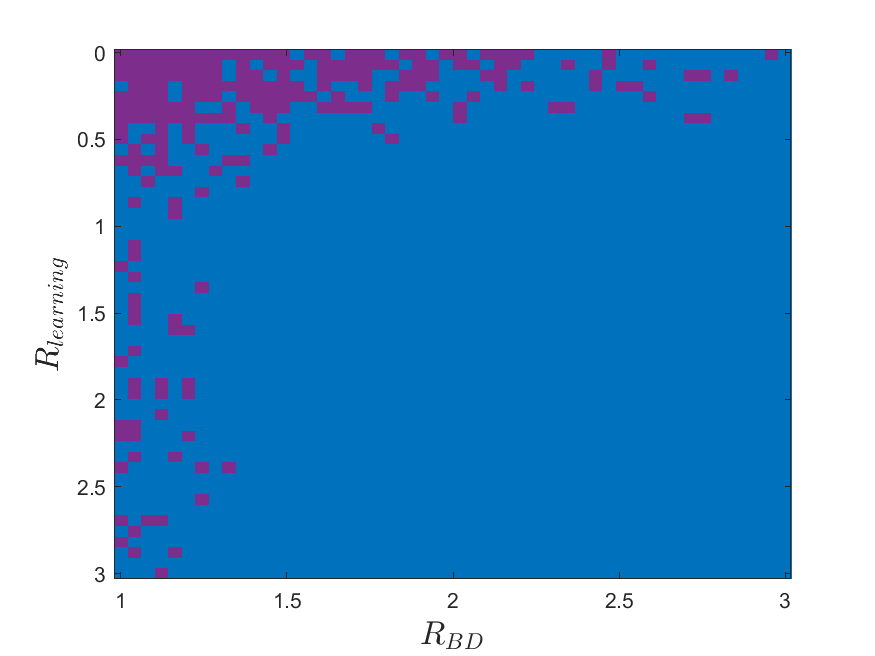}
\end{minipage}
\begin{minipage}{0.32\textwidth}
\centering
\includegraphics[scale=0.38]{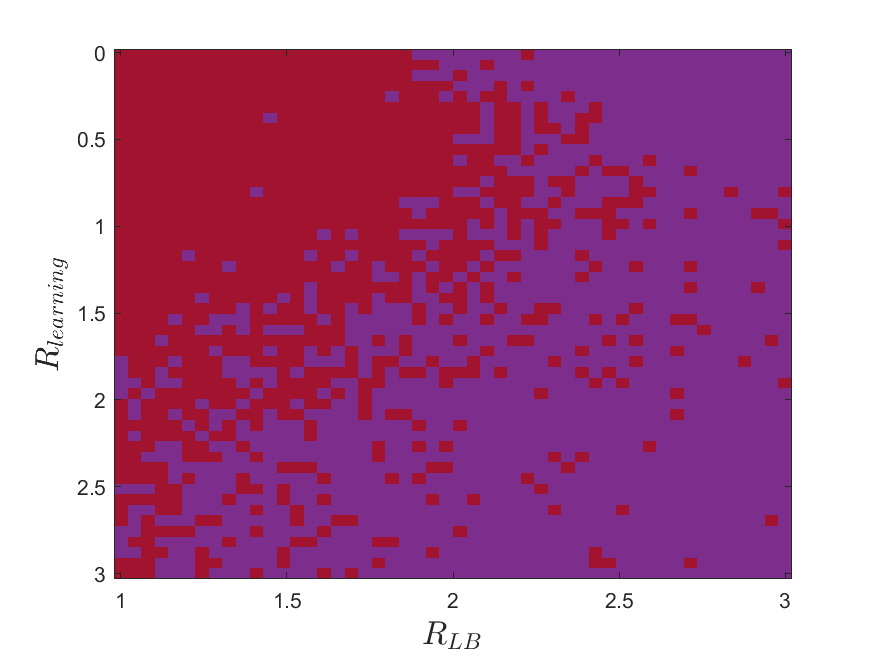}
\end{minipage}
\caption{Winning agent types as a function of relative replication cost and learning cost for mixed populations consisting of 100 Lamarckian and 100 Darwinian agents (left panel); 100 Baldwinian and 100 Darwinian agents (middle panel); 100 Lamarckian and 100 Baldwinian agents (right panel).}
\label{comp_E}
\end{figure}
It is found that learning gives a definite advantage at low learning cost; even when learning agents have a high relative replication cost compared to non-learning agents. Also note that the Baldwinian agent type has an advantage over the Lamarckian agent type at low learning cost and $R_{LB}>1.8$, which is further demonstrated in Figure \ref{fig:threeway}.

\subsection{Three-way agent population competition}\label{Three-way}

We now investigate a system similar to the one in the previous section, but with all three agent types in competition.

\begin{figure}[H]
\centering
\advance\leftskip-1.25cm
\includegraphics[scale=0.44]{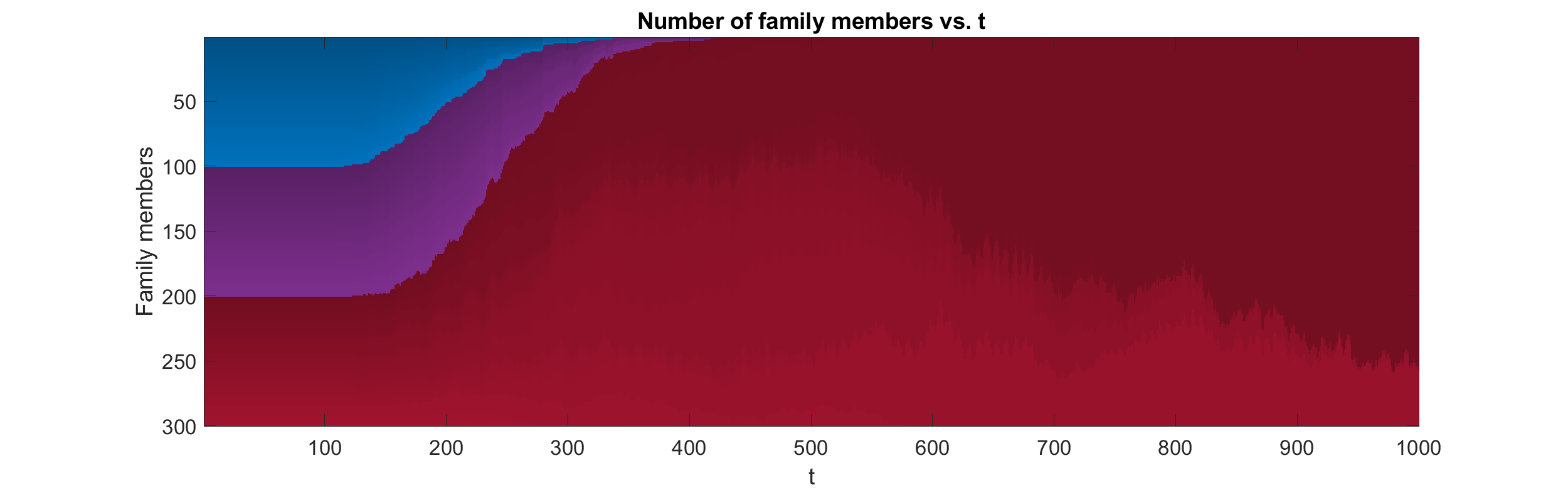}
\caption{Family dynamics of a mixed initial population consisting of 100 Darwinian agents (blue spectrum), 100 Baldwinian agents (purple spectrum), and 100 Lamarckian agents (red spectrum). For the used parameters and initial conditions, the Lamarckian agent type fills the population at $t\approx 400$.}
\label{comp_fam_dyn}
\end{figure}

\begin{figure}[H]
\centering
\includegraphics[scale=0.60]{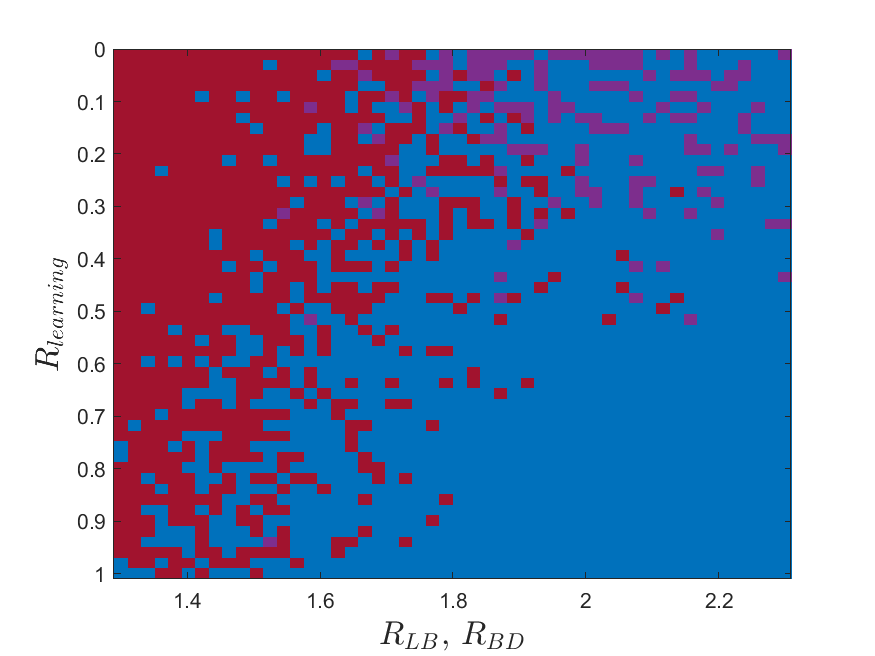}
\caption{Investigation of three-way competition with an initial population consisting of 100 agents of each agent type. Here, we have used $R_{learning}=\{0.01,...,  1\}$ and $R_{LB}=R_{LB}=\{1.3,..., 2.3\}$.}
\label{fig:threeway}
\end{figure}

For this, we limit our investigations to cases where $R^L_{replication} \leq R^B_{replication} \leq R^D_{replication}$; more specifically, we define $R_{LB}=R_{BD}$, e.g. for the rightmost column of pixels in Figure \ref{fig:threeway} we define $R_{LB}=R_{BD}=2$, meaning $R^L_{replication}=2R^B_{replication}=4R^D_{replication}$.

The reasoning behind this choice is as follows: In biology, the cost of replication commonly increases with the complexity and amount of the wetware that is replicated. We argue that Baldwinian organisms are typically more complex than simple Darwinian organisms, as the capability of learning requires additional wetware. It is then reasonable that Lamarckian organisms would require even more wetware to also encode the learned phenotypic changes.

It can be seen that all three agent types have regions in this parameter space where their algorithms are advantageous. The Lamarckian agent type predominantly wins at $R_{LB}=R_{LB}<1.8$ The Darwinian agent type wins at high learning costs and high relative replication costs, while the Baldwinian agent type frequently wins at $R_{LB}=R_{LB}>1.8$ and $R_{learning}<0.5$, which corresponds with \autoref{comp_E}.

\begin{figure}[H]
\centering
\includegraphics[scale=0.55]{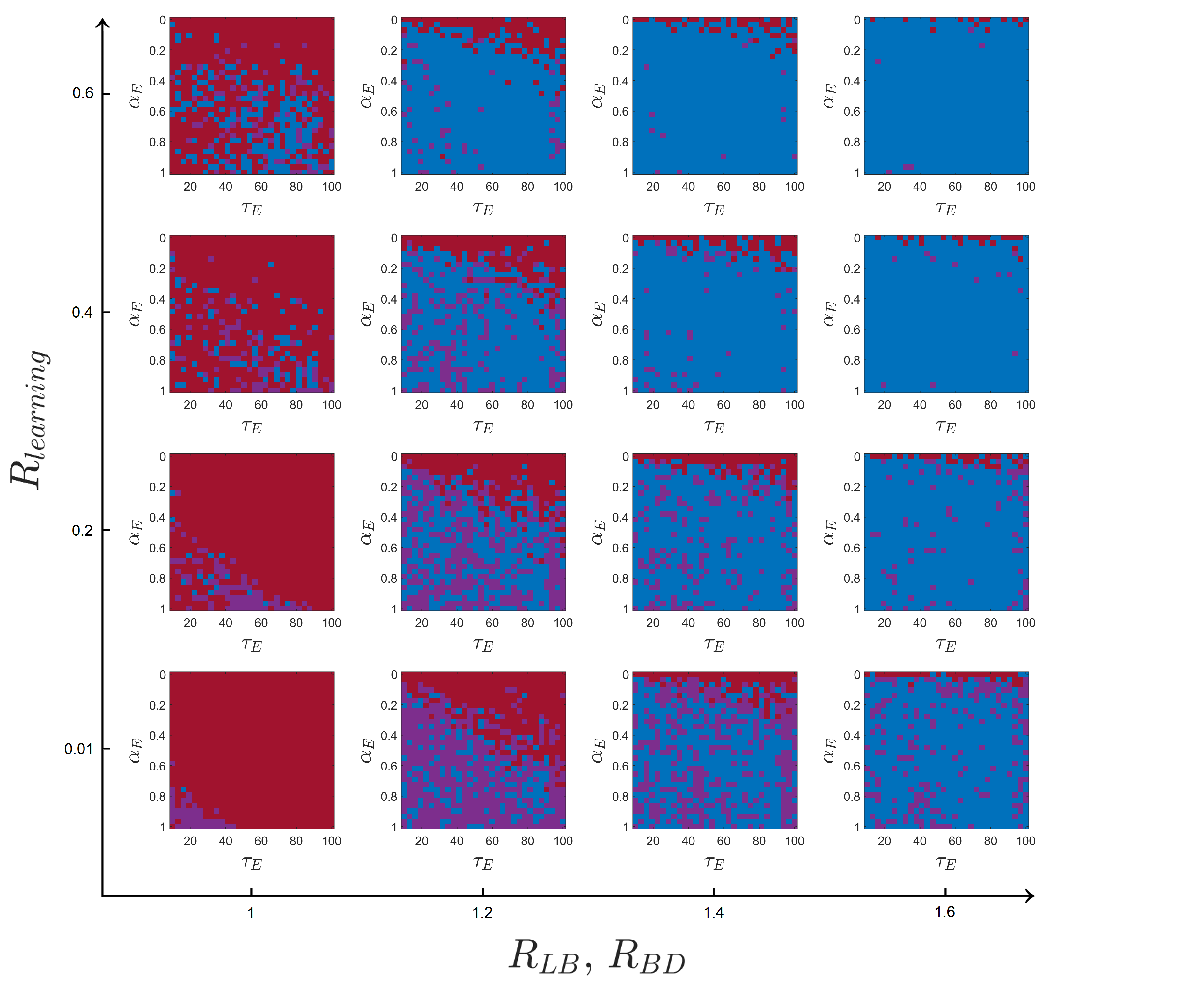}
\caption{Each subplot shows the winning agent types in dynamic environments described by pairs of values for the environment mutation period ($\tau_E$) and the environment mutation rate ($\alpha_E$). Furthermore, these subplots are made for 16 different pairs of values for the relative replication cost ($R_{LB}$ and $R_{BD}$, constrained by demanding $R_{LB}=R_{BD}$) and learning cost ($R_{learning}$).}
\label{fig:4x4threeway}
\end{figure}

So far, the competition dynamics have only been investigated in constant environments. We now allow the environment to change, as described in Section \ref{sec: E(t)}, while also varying the learning cost and relative replication cost, thereby forming a multiaxial plot of the winning agent type as can be seen on \autoref{fig:4x4threeway}.

Much information of the model dynamics can be extracted from \autoref{fig:4x4threeway}. At equal replication cost ($R_{LB}=R_{BD}=1$), the Lamarckian agent dominates in the dynamic environments. At $R_{LB}=R_{BD}\geq1.2$ the Darwinian and Baldwinian agent types dominate in very dynamic environments, i.e., at high $\alpha_E$ and low $\tau_E$. The fraction of Darwinian to Baldwinian winners depend heavily on the learning cost, with only a small dependence on $\alpha_E$ and $\tau_E$. A small $\alpha_E$ always benefits the Lamarckian agent type. This indicates that the curiosity learning algorithm is efficient at adapting phenotypes to minor changes in the environment. Finally, at both high learning cost and high relative replication cost the Darwinian agent type dominates in the dynamic environments, though the Lamarckian agent type still frequently wins in a constant and near-constant environment. This goes against our hypothesis - supported by our initial findings - of the Lamarckian agent type being the most fit in a dynamic environment; at least there is a large dependence on the relative replication cost and the specific nature of the dynamic environment.



\section{Discussion}

Though the main focus during the design of the model has been conceptual simplicity, the process of formulating the model into code has involved the introduction of many variables. This has made the set of possible initial conditions too large to make a complete investigation of the possible dynamics of the system. The study of the Baldwinian agent type was mostly included to satisfy a curiosity of the role of learning in biological systems, but the small genome size used in simulation is far from biological genomes which are on the scale of $10^5-10^{12}$ base pairs \cite{phillips2012physical}, and as mentioned in Section \ref{section_Darwinian_agents} the simple genotype-phenotype mapping also deviates from the complex mapping in biological systems \cite{pigliucci2010genotype}. As shown in Figure \ref{N=1000 L=30}, the Lamarckian population is less likely to down-regulate curiosity when having large genomes which limit evolutionary optimization at high fitness of the current Lamarckian algorithm. Furthermore, the curiosity algorithm has a maximal learning rate of a single bit per $t$-step, where the Darwinian algorithm instead is maximally able to change $\alpha L$ bits per generation. This proportionality gives the Darwinian algorithm a definite advantage at very large genome sizes. The amount of inherited information of even the simplest biological organisms capable of learning surely is much greater then the amount of genetic information carried by the implemented agents in all simulations, so the currently implemented curiosity algorithm is unequipped for optimizing such systems. This is all specific to the learning algorithm, and therefore no generic conclusions can be stated about the general dynamics of other learning algorithms for systems of high information.

The reasoning behind the design of the learning algorithm used by the Lamarckian and Baldwinian agents has been conceptional and computational simplicity. A different learning algorithm would undoubtedly yield different dynamics, which subsequently limits the utility of the specific findings outlined in the Results section for general systems. Though this must be the case, there are aspects of the found results which may be generic. Particularly, the advantage of agents having the ability to learn during its lifetime while in an environment which changes at a faster rate than the replication rate of the agents. The ability of the Lamarckian agent to transmit such learned changes during replication has yielded the fastest rise in fitness for most of the investigated parameter conditions.

As the simulations fall under the category of mathematical optimization, it is relevant to compare and discuss the computational cost of simulations for each agent type. The main computational cost is due replication of agents and the calculation of the fitness of each agent. In a constant environment the Darwinian agents have constant fitness during their lifetime, their fitness is therefore only calculated once; at birth. The Lamarckian and Baldwinian agent types change their fitness during their lifetime at each learning event, and their new fitness must be calculated whenever they learn adding a scaling factor of $\Bar{c}$ to the computation cost. In terms of the model-specific parameters, it is mostly the maximum curiosity $C$ and the replication threshold $R$ which determine the average computational cost of each $t$-step. The stated runtimes on the figures in Section \ref{sec:comparison} show a high cost of the Lamarckian algorithm, but for most parameters the Lamarckian populations quickly reach a high fitness, and computational cost could be lowered by gauging the runtime needed for reaching the desired fitness, or simply terminating the simulation when a desired fitness has been reached.
 
Other fitness functions could also easily be implemented in the model for investigation. The simple linear fitness landscape of the environment used in this study could be exchanged with, e.g., a more complex NK model.

Further studies of Lamarckian and Baldwinian evolution and optimization could implement more sophisticated learning algorithms based on artificial neural networks or simulated annealing in more complex environments, perhaps with defined temporal or spacial patterns.

A system consisting of agents utilizing neural networks to distinguish food from poison in an environment of dynamically changing set of materials has been studied by Sasaki and Tokoro (1999) \cite{sasaki1999evolving}. By varying the heritability of acquired characteristics, the authors find the Darwinian population to have the highest level of adaption in dynamic environments. Our model shows similar results (\autoref{fig:4x4threeway}), through not generally, as each agent type may perform best in dynamic environments depending on the model parameter values.

\newpage
\section{Conclusions}

Our work successfully implements the Darwinian, Lamarckian and Baldwinian algorithms in a single agent based modelling framework, and all three algorithms exhibit evolutionary optimization. Pure populations of all agent types exhibit a region of exponential growth in agent replications if the population is not yet filled with replicating agents (Figure \ref{replication_dyn} left panel). When all agents in the population is able to replicate, the replication rate instead increases slowly with population mean fitness (Figure \ref{replication_dyn} right panel).

Fitness dynamics generally deviate from standard optimization processes in $log(t)$-plots at low and high fitness due to the fitness dependency of replication rates (Figure \ref{D f dyn}). The speed of evolutionary optimization for pure Darwinian populations is inversely proportional to the replication threshold, meaning a high replication rate benefits the Darwinian agent type (Figure \ref{N=1000 R=5}). Optimization speeds for pure Lamarckian agent populations are proportional to curiosity (learning rate), though the maximal fitness is inversely proportional to curiosity (Figure \ref{N=1000 C=0.01}). At large genome sizes pure Lamarckian populations are less likely to reach a final state of zero curiosity via down-regulation of the curiosity gene (Figure \ref{N=1000 L=30}). The evolutionary optimization speeds of pure Baldwinian populations are initially higher than pure Darwinian populations, but become lower at high fitness as the curiosity learning algorithm becomes less efficient since the resources expended on learning instead inhibits the optimization of Baldwinian populations via the Darwinian algorithm due to the lower replication rate compared to the Darwinian agent type (Figure \ref{N=1000}). For high maximum curiosity, optimization speeds of pure Baldwinian populations are slower than those of pure Lamarckian populations due to the loss of useful phenotypic changes at replication, but at low curiosity the Baldwinian dynamics coincide with the dynamics of Darwinian populations (Figure \ref{N=1000 C=0.01}). 

Both Lamarckian and Baldwinian populations reach higher phenotypic fitness than the Darwinian population in a temporally dependent environment which scrambles at a periodicity on the same timescale as the replication rate (Figure \ref{temporalE}).

Mixed populations simulations allowing for direct competition between agent types (pairwise and three-way) show populations drift toward homogeneity. These simulations show regions in parameter space that benefit the agent types. 

\newpage

\bibliographystyle{plain.bst}
\bibliography{bibliography.bib}

\newpage
\addcontentsline{toc}{section}{Appendix A}

\end{document}